\documentclass[trackchanges,twocolumn]{aastex701}

\usepackage{txfonts}
\usepackage{url}
\usepackage{hyperref}
\usepackage{orcidlink}
\usepackage{graphicx}
\usepackage{cleveref}
\usepackage{txfonts}
\usepackage{subfig}

\begin{document}

\title{POLAMI Multi-Wavelength Polarization Study of AGN Jets: A Millimeter-Optical Comparison}

\author[orcid=0000-0003-1117-2863, sname='Casadio']{Carolina~Casadio}
\affiliation{Institute of Astrophysics, Foundation for Research and Technology-Hellas, GR-70013 Heraklion, Greece}
\affiliation{Department of Physics, University of Crete, GR-70013, Heraklion, Greece}
\email[show]{ccasadio@ia.forth.gr}  

\author[orcid=0000-0003-0611-5784, sname='Blinov']{Dmitry~Blinov} 
\affiliation{Institute of Astrophysics, Foundation for Research and Technology-Hellas, GR-70013 Heraklion, Greece}
\affiliation{Department of Physics, University of Crete, GR-70013, Heraklion, Greece}
\email[show]{blinov@physics.uoc.gr}  

\author[orcid=0000-0002-3777-6182,sname='Agudo']{Iván~Agudo}
\affiliation{Instituto de Astrof\'{\i}sica de Andaluc\'{\i}a-CSIC, Glorieta de la Astronom\'{\i}a s/n, 18008 Granada, Spain}
\email{iagudo@iaa.es}

\author[orcid=0000-0002-3777-6182, sname='Myserlis']{Ioannis~Myserlis}
\affiliation{Institut de Radioastronomie Millimétrique, Avenida Divina Pastora, 7, Local 20, E-18012 Granada, Spain}
\email{imyserlis@iram.es}

\author[orcid=0009-0006-5292-6974,sname='Thum']{Clemens~Thum}
\affiliation{Institut de Radioastronomie Millimétrique, Avenida Divina Pastora, 7, Local 20, E-18012 Granada, Spain}
\email{thum@iram.es}

\author[orcid=0000-0001-6158-1708, sname='Jorstad']{Svetlana~Jorstad}
\affiliation{Institute for Astrophysical Research, Boston University, 725 Commonwealth Avenue, Boston, MA 02215, USA}
\affiliation{Saint Petersburg State University, 7/9 Universitetskaya nab., St. Petersburg, 199034, Russia}
\email{jorstad@bu.edu}

\author[orcid=0000-0001-7396-3332, sname='Marscher']{Alan~Marscher}
\affiliation{Institute for Astrophysical Research, Boston University, 725 Commonwealth Avenue, Boston, MA 02215, USA}
\email{marscher@bu.edu}

\author[orcid=0000-0001-9826-1759]{Haocheng~Zhang}
\affiliation{University of Maryland, Baltimore County, Baltimore, MD 21250, USA}
\affiliation{NASA Goddard Space Flight Center, Greenbelt, MD 20771, USA}
\email{phitlip2007@gmail.com}

\author[orcid=0000-0002-4131-655X, sname='Escudero']{Juan~Escudero~Pedrosa}
\affiliation{Center for Astrophysics | Harvard \& Smithsonian, Cambridge, MA 02138, USA}
\affiliation{Instituto de Astrof\'{\i}sica de Andaluc\'{\i}a-CSIC, Glorieta de la Astronom\'{\i}a s/n, 18008 Granada, Spain}
\email{juan.escuderopedrosa@cfa.harvard.edu}

\author[orcid=0000-0002-9998-5238]{Diego~\'Alvarez-Ortega}
\affiliation{Institute of Astrophysics, Foundation for Research and Technology-Hellas, GR-70013 Heraklion, Greece}
\affiliation{Department of Physics, University of Crete, GR-70013, Heraklion, Greece}
\email{dalvarez@physics.uoc.gr}

\author[orcid=0000-0001-6314-0690]{Zachary~R.~Weaver}
\affiliation{Institute for Astrophysical Research, Boston University, 725 Commonwealth Avenue, Boston, MA 02215, USA}
\email{zweaver@bu.edu}

\author[orcid=0000-0003-1134-7352]{Manasvita~Joshi}
\affiliation{FAS Research Computing, Harvard University, 38 Oxford Street, Cambridge, MA 02138}
\email{manasvitajoshi@g.harvard.edu}

\author[orcid=0000-0002-3375-3397]{Callum~McCall}
\affiliation{Astrophysics Research Institute, Liverpool John Moores University, Liverpool Science Park IC2, 146 Brownlow Hill, Liverpool L3 5RF, UK}
\email{callumlmccall@gmail.com}

\author[orcid=0000-0002-3375-3397]{Helen~Jermak}
\affiliation{Astrophysics Research Institute, Liverpool John Moores University, Liverpool Science Park IC2, 146 Brownlow Hill, Liverpool L3 5RF, UK}
\email{h.e.jermak@ljmu.ac.uk}

\author[orcid=0000-0001-8397-5759]{Iain~A.~Steele}
\affiliation{Astrophysics Research Institute, Liverpool John Moores University, Liverpool Science Park IC2, 146 Brownlow Hill, Liverpool L3 5RF, UK}
\email{i.a.steele@ljmu.ac.uk}

\author[orcid=0000-0002-7262-6710]{George~A.~Borman}
\affiliation{Crimean Astrophysical Observatory RAS, P/O Nauchny, 298409}
\email{borman.ga@gmail.com}

\author[orcid=0000-0002-3953-6676]{Tatiana~S.~Grishina}
\affiliation{Saint Petersburg State University, 7/9 Universitetskaya nab., St. Petersburg, 199034, Russia}
\email{t.s.grishina@spbu.ru}

\author[orcid=0000-0002-2471-6500]{Elena~G.~Larionova}
\affiliation{Saint Petersburg State University, 7/9 Universitetskaya nab., St. Petersburg, 199034, Russia}
\email{sung2v@mail.ru}

\author[orcid=0000-0002-9407-7804]{Daria~A.~Morozova}
\affiliation{Saint Petersburg State University, 7/9 Universitetskaya nab., St. Petersburg, 199034, Russia}
\email{d.morozova@spbu.ru}

\author[orcid=0000-0003-4147-3851]{Sergey~S.~Savchenko}
\affiliation{Saint Petersburg State University, 7/9 Universitetskaya nab., St. Petersburg, 199034, Russia}
\affiliation{Pulkovo Observatory, St.Petersburg, 196140, Russia}
\email{s.s.savchenko@spbu.ru}

\author[orcid=0000-0002-4218-0148]{Ivan~S.~Troitskiy}
\affiliation{Saint Petersburg State University, 7/9 Universitetskaya nab., St. Petersburg, 199034, Russia}
\email{i.troitsky@spbu.ru}

\author[orcid=0000-0002-9907-9876]{Yulia~V.~Troitskaya}
\affiliation{Saint Petersburg State University, 7/9 Universitetskaya nab., St. Petersburg, 199034, Russia}
\email{y.troitskaya@spbu.ru}

\author[orcid=0000-0002-8293-0214]{Andrey~A.~Vasilyev}
\affiliation{Saint Petersburg State University, 7/9 Universitetskaya nab., St. Petersburg, 199034, Russia}
\email{andrey.vasilyev@spbu.ru}

\begin{abstract}

Millimeter-band polarimetry offers a powerful probe of AGN jets, accessing regions less affected by opacity and Faraday rotation than at longer radio wavelengths. As part of the POLAMI program, we have conducted 14 years of 1 mm and 3 mm polarization monitoring of a sample of gamma-ray–bright blazars with the IRAM 30-m telescope, complemented here with long-term optical polarimetric observations from multiple facilities.
We aim to test whether current models of parsec-scale jet physics are consistent with observed multi-band polarization behavior. Using a Bayesian framework, we derive intrinsic mean flux densities and modulation indices for total flux and fractional polarization, and characterize EVPA variability using circular statistics. We then examine how these quantities reflecting variability properties across millimeter and optical bands relate to synchrotron peak frequency, jet orientation, and radio/gamma-ray luminosities.
BL Lac objects exhibit, on average, higher fractional polarization and lower EVPA variability than FSRQs at all wavelengths. Fractional polarization increases with frequency, consistent with increasingly ordered magnetic fields at shorter wavelengths. BL Lacs also show more frequent alignment of EVPAs between optical and millimeter bands, whereas FSRQs display weaker coherence. EVPA variability correlates positively with radio and gamma-ray luminosities and negatively with synchrotron peak frequency, most strongly in the optical. We further find a positive correlation between EVPA spread and fractional polarization variability, suggesting a direct link between magnetic-field structure and polarization dynamics.

\end{abstract}



\section{Introduction}\label{sec:intr}

Relativistic jets in active galactic nuclei (AGN) are considered Poynting-flux dominated at their origin, with strong magnetic fields that influence the jet flow dynamics and particle acceleration \citep{Blandford1977, McKinney2006}. Linear polarization studies at short millimeter wavelengths and optical frequencies, where the emission is believed to be optically thin and, therefore, free of opacity effects, are fundamental to investigating the behavior and, consequently, the role of magnetic fields in jets. 

Blazars are a subset of radio-loud AGN characterized by powerful relativistic jets that are closely aligned with our line of sight \cite[e.g.,][]{Padovani2017}. They are typically classified into flat-spectrum radio quasars (FSRQs) if they display strong, broad, and narrow emission lines in their optical spectra, and BL Lacertae objects (BL Lacs) if they show weak or no emission lines \citep{UrryPadovani1995}. The bulk jet motion, coupled with the small viewing angle, results in Doppler boosting, which amplifies the jet's brightness and shortens observed variability timescales.

While the millimeter wave emission in blazars is expected to be dominated by synchrotron radiation from the relativistic jet, optical emission can have multiple production mechanisms (synchrotron, thermal emission or recombination lines emission) and locations (accretion disk, broad line region, or jet) \citep{Podjed_2024}. 
Linear polarization studies can help distinguish between all these possible scenarios. Synchrotron emission from beamed jets is expected to exhibit a variable degree of polarization, ranging from a few to several tens of percent. A strong contribution from either the non-beamed synchrotron components or the thermal components, like the accretion disk, broad line region (BLR), or torus, can lower the fractional polarization to less than a few percent \citep{Bottcher2017}. At the same time, these components are expected to reduce the variability in the combined emission’s total flux and polarization.  

\cite{Agudo2018I} presented the POLAMI (Polarimetric Monitoring of AGN at Millimetre Wavelengths) program for the study of the polarimetric properties of a sample of radio and gamma-ray bright blazars and radio-galaxies with the IRAM 30-m telescope at 3.5 and 1.3 mm, and bi-weekly (on average) cadence. 
New important results on polarization variability in POLAMI sources were presented in \citep{Agudo2018III}, namely: i) the fractional polarization at 1~mm is found to be higher, more variable and to vary faster than at 3~mm; ii) the linear polarization angle is in general highly variable and the 1~mm variations seem to be faster than those at 3~mm. Faster variability and larger amplitudes at 1~mm than at 3~mm are also seen in total flux density. 

\cite{Agudo2018III} confirmed that opacity effects are not the dominant cause behind the variability observed at millimeter wavelengths. In general, sources are optically thin at both 1 and 3 millimeters, becoming optically thick at 3~mm only during certain prominent flares and for periods shorter than the flaring state duration. POLAMI's results support a multi-zone model in which 1~mm emission originates in more compact regions with better-ordered magnetic fields than at 3~mm, as predicted by the turbulent extreme multi-zone (TEMZ) model, where turbulence drives the variability of jet emission \citep{Marscher2014}. A certain degree of magnetic-field disorder is required to explain the relatively low fractional polarization observed, which is consistently far below the theoretical maximum \citep[$\sim$70\%, ][]{Pacholczyk1970} for synchrotron radiation in a uniform field. However, the frequent alignment between the linear polarization angle at optical frequencies and those of the radio core \citep{Marscher&Jorstad} or stationary jet features \citep{Sasada2018} indicates that an ordered magnetic-field component must also be present. Pure turbulence alone cannot even explain the systematic rotations observed in the optical linear polarization vectors of some blazars \citep{Blinov2018, Blinov2021}.

This work presents a systematic comparative study of optical and millimeter-wavelength polarimetric observations of a sample of blazars. Our goal is to address the long-standing questions concerning the origin of optical emission in blazars and the intrinsic physical conditions that differentiate them. For this study, we make use of the long-term POLAMI program for the millimeter data and several optical monitoring programs for the optical data. While this paper concentrates on integrated quantities, a complementary time-dependent variability analysis is currently in preparation.

%%%%%%%%%%%%%%%%%%%%%%%%%%%%%%%%%%%%%%%%%%%%%%%%%%
\section{Data and Methods}\label{sec:data}
The priority sample of the POLAMI program, considered in this study, includes 34 sources in total: 21 FSRQs, 11 BLLacs, and 2 Radio Galaxies. The list of sources with the associated class is reported in Tab.~\ref{tab:sources_list}. POLAMI observations at 1.3~mm (230~GHz) and 3.5~mm (86.24~GHz) were regularly (average sampling $\sim$20 days) carried out at the IRAM 30~m Telescope in Pico Veleta since 2006 (2009 at 1~mm). Observation strategy and data calibration are described in detail \cite{Agudo2018I}. 
The millimeter radio data were complemented by optical photo-polarimetric observations from the following telescopes: (1) the 2.2~m telescope of the Calar Alto Observatory (Almería, Spain)\footnote{Observations from the Monitoring AGN with the Calar Alto Telescopes (MAPCAT) program; see~\cite{Agudo2012IJMPS...8..299A}};  (2) the 2~m Liverpool Telescope of the Observatorio del Roque de Los Muchachos (Canary Islands, Spain); (3) the 1.83~m Perkins Telescope of Lowell Observatory (Flagstaff, USA)\footnote{from 2019 the Perkins telescope is Perkins Telescope Observatory (PTO) owned by Boston University}; (4) the 1.54~m and 2.3~m telescopes of Steward Observatory (Mt.Bigelow and Kitt Peak, USA) \footnote{\url{https://james.as.arizona.edu/~psmith/Fermi/}}; (5) the 70~cm AZT-8 Telescope of the Crimean Astrophysical Observatory; (6) the 40~cm LX-200 Telescope of St. Petersburg State University (St. Petersburg, Russia). Additional data were obtained from the publicly available RoboPol \citep{Blinov2021MNRAS.501.3715B} and Kanata \citep{Itoh2016} monitoring programs, using (7) the 1.3~m telescope of the Skinakas Observatory (Crete, Greece), and (8) the 1.5~m Kanata Telescope at the Higashi-Hiroshima Observatory (Japan). Most of the optical polarimetric data were obtained in the R-band, with a few exceptions. The Steward Observatory data were collected using the SPOL spectropolarimeter over the 5000–7000~{\AA} range. At the LX-200 telescope, polarimetry was conducted in white light, with an effective bandpass approximately matching the R-band. For three sources observed in the Kanata program, V-band data were used. In the case of the Liverpool Telescope, RINGO3 polarimetric measurements were derived by averaging the ‘f’ and ‘e’ bands \citep{Arnold2012}. Although not all measurements were made strictly in the R-band, the effective wavelengths are sufficiently close, and optical polarization parameters typically vary only weakly and irregularly with wavelength \citep{Marscher&Jorstad}. Therefore, these data were treated as equivalent to R-band measurements and included in the analysis. For the total optical flux density, we used only photometry obtained in the R-band, which was corrected for the Galactic extinction using data from \cite{Schlafly2011}. The total flux and fractional polarization were not corrected for host-galaxy emission, as such contributions are negligible for most sources.

POLAMI data at 3 mm and 1~mm cover the periods from October 2006 to October 2021 and from December 2009 to October 2021, respectively, corresponding to average total coverages of roughly 14 years (3~mm) and 11 years (1~mm). Optical data are collected from April 2005 to September 2021 and cover, on average, 15 years. The average data sampling per source at 3~mm goes from 17 to 47 days and from 24 to 146 days at 1~mm. At optical frequencies, the average data sampling per source ranges from intraday scales to 18 days. 

We inspected data at all three frequencies and discarded electric vector position angle (EVPA) measurements when $p/\sigma_p < 2.5$, where $p$ is the fractional polarization and $\sigma_p$ its uncertainty. After excluding unreliable EVPAs, if a source is left with less than 20 EVPA measurements at any band, we did not compute the average EVPA value at that band. 

To determine the mean flux density and a measure of its variability, we employed the Bayesian approach introduced by \cite{Richards2011}. The intrinsic modulation index quantifies the amplitude of variability in the flux density of a source, independent of observational uncertainties and sampling effects, and is defined as the ratio of the intrinsic standard deviation of the flux density distribution to its intrinsic mean, $m_{\rm fl} = \sigma_S / S_0$, where “intrinsic” refers to the variability that would be observed with perfect sampling and no measurement errors. The Bayesian method from \cite{Richards2011} allows for the simultaneous estimation of the maximum likelihood values of both $S_0$ and $m_{\rm fl}$, providing information about the source’s intrinsic flux density variability and average flux, deconvolved from both observational errors and the effects of finite sampling.

The approach of \cite{Richards2011} was extended to polarization measurements in \cite{Blinov2016} to characterize the intrinsic average fractional polarization, $p_0$, and the intrinsic modulation index of the fractional polarization, $m_{\rm pol}$. The intrinsic average fractional polarization, $p_0$, represents the mean polarization fraction of a source, corrected for observational biases and sampling effects, while $m_{\rm pol}$ quantifies the amplitude of variability in the fractional polarization, independent of measurement uncertainties and sampling limitations, and is defined as the ratio of the intrinsic standard deviation to the intrinsic mean, $m_{\rm pol} = \sigma_{p} / p_0$. A key advantage of this technique is that, in addition to accounting for measurement uncertainties and sampling effects, it also corrects for polarization bias \citep[e.g.,][]{Vaillancourt2006}.

We modeled the observed dependencies using linear least-squares fits. The significance of the fits was tested by applying the same statistical analysis to bootstrapped data. Bootstrapping was performed by resampling the original dataset both with and without replacement. The procedure was repeated 500 times, and the mean and standard deviation of the fitted slopes were compared with those obtained from the original data. We found that the bootstrapping analysis did not change the significance of the correlations; therefore, we consider the analyzed dataset to be a good representation of the parent population.    

\begin{table}
	\centering
	\caption{List of sources}
	\label{tab:sources_list}
     \begin{tabular}{lccc} % 5 columns, alignment for each
		\hline
		J2000 name & Common name & Opt. Class & Ref. \\
		\hline
J0222+4302 &  3C66a      & B & R14 \\
J0238+1636 &  AO0235+16  & B & R14 \\
J0319+4130 &  3C84       & RG & VCV10\\
J0339-0146 &  CTA26      & Q & R14 \\
J0418+3801 &  3C111      & RG & VCV10 \\
J0423-0120 &  PKS0420-01 & Q & R14 \\
J0530+1331 &  PKS0528+13 & Q & R14 \\
J0721+7120 &  S50716+71  & B & R14\\
J0738+1742 &  PKS0735+17 & B & R14\\
J0830+2410 &  OJ248      & Q & R14\\
J0831+0429 &  OJ49       & B & R14\\
J0841+7053 &  Q0836+71   & Q & R14 \\ 
J0854+2006 &  OJ287      & B & R14 \\ 
J0958+6533 &  S40954+65  & B & R14  \\
J1058+0133 &  PKS1055+01 & Q & R14 \\
J1104+3812 &  MKN421     & B & R14 \\
J1130-1449 &  PKS1127-14 & Q & VCV10\\
J1159+2914 &  4C 29.45   & Q & R14 \\
J1221+2813 &  WComae     & B & R14 \\ 
J1224+2122 &  PKS1222+21 & Q & R14 \\
J1229+0203 &  3C273      & Q & R14 \\ 
J1256-0547 &  3C279      & Q & R14 \\
J1310+3220 &  B21308+32  & Q & R14 \\
J1408-0752 &  PKS1406-07 & Q & R14 \\
J1512-0905 &  PKS1510-08 & Q & R14 \\ 
J1613+3412 &  DA 406     & Q & R14 \\
J1635+3808 &  4C 38.41   & Q & R14 \\ 
J1642+3948 &  3C345      & Q & R14 \\
J1733-1304 &  Q1730-13   & Q & R14 \\
J1751+0939 &  OT081      & B & R14 \\
J2202+4216 &  BLLac      & B & R14 \\
J2225-0457 &  3C446      & Q & R14 \\
J2232+1143 &  CTA102     & Q & R14 \\
J2253+1608 &  3C454.3    & Q & R14 \\
\hline 
 \end{tabular}
 \begin{minipage}{0.47\textwidth}
 Notes: Optical classes (B: BL Lac, RG: radio galaxy, Q: quasar) are taken from \cite{Richards2014} (R14) and \cite{VCV2010} (VCV10).
 \end{minipage}
\end{table}

%%%%%%%%%%%%%%%%%%%%%%%%%%%%%%%%%%%%%%%%%%%%%%%%%%
\section{Results}

\subsection{Average Quantities: EVPA, jet PA, and $p_{0}$}\label{sec:evpa_pa}

\begin{figure}
\centering
\includegraphics[width=0.37\textwidth]{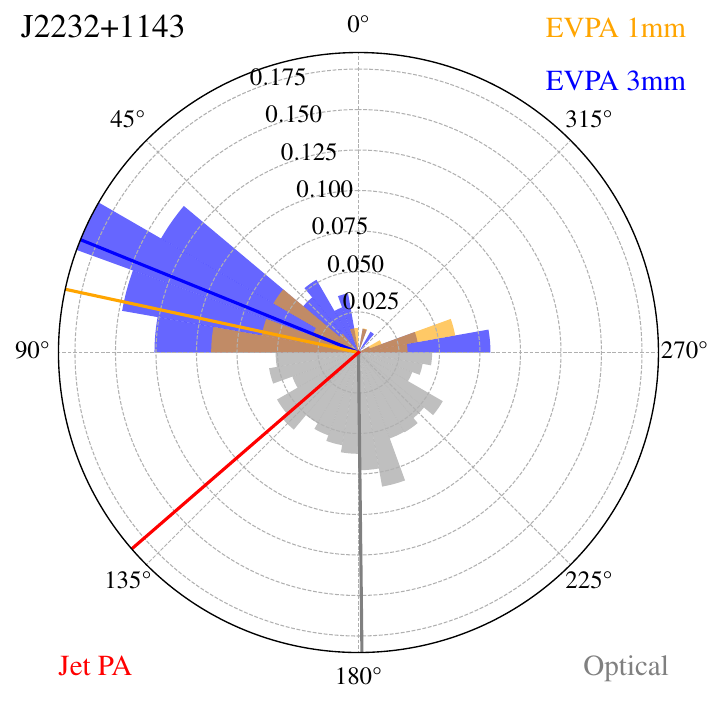}
\caption{Millimeter (upper hemisphere) and optical (lower hemisphere) EVPA distribution in J2232+1143. EVPA distributions at 3~mm (blue), 1~mm (yellow), and optical (grey) wavelengths are overlaid by vectors of the same color representing the mean value of the respective distribution. The red line marks the jet PA. The different circles mark the counts as a fraction of the total. Corresponding distributions for all other sources in the sample are provided in Appendix~\ref{app:A}.}
\label{fig:cir_hist_one}
% ./polami/5_circular_hist/plot_hist_opt_mm.py
\end{figure}

We test the uniformity of the EVPA distribution using a Kolmogorov–Smirnov (K–S) test. The null hypothesis of EVPAs following a uniform distribution is rejected if the likelihood of data occurring under the null hypothesis (p-value) is lower than 0.05.
Based on the K–S test, the majority of sources (68\%) exhibit non-uniform distributions in millimeter and optical EVPAs. For sources where the EVPA is not uniformly distributed, we compute the mean value and its associated uncertainty using circular statistics. We then compare these mean EVPA values with the jet position angles (PAs) reported by \cite{Weaver2022}. It is important to note that a large fraction of sources in this study (38\%) have variable jet position angle, based on \cite{Weaver2022} classification.
We visualize the distribution of EVPAs at the different bands in circular histograms, as shown in Fig.~\ref{fig:cir_hist_one}. The red vector represents the jet PA, while blue and yellow vectors indicate the mean of 3 and 1~mm EVPAs, respectively, and are overlaid on the plot if EVPAs do not follow a uniform distribution. In some sources, the optical and millimeter EVPAs tend to be stable around the mean, with $>$70$\%$ of the measurements confined within $\pm$25$^{\circ}$ of the mean. In general, the stability around a mean orientation is observed more often in 3~mm EVPAs (6 sources: J0854+2006, J0958+6533, J1224+2122, J1229+0203, J1256-0547, J2202+4216) than at 1~mm (2 sources: J0854+2006 and J2202+4216) and optical (4 sources). The paucity of sources at 1~mm may reflect both limited data and stronger variability at this wavelength.
%Sources exhibiting such behavior at either 1 or 3~mm are J0854+2006, J0958+6533, J1224+2122, J1229+0203, J1256-0547, and J2202+4216. Only in the two BL Lac objects, J0854+2006 and J2202+421, both 1 and 3~mm EVPAs show stability. At the optical band, only two sources, J1221+2813 and J1224+2122, have EVPAs stable around the mean. 

%%%%% Table %%%%%
 \begin{table*}[]
 \centering
 \caption{Differences in the mean EVPA between bands (columns 3–5) and between the mean EVPA in the three bands and the jet position angle (PA) (columns 6–8). Sources for which the EVPA distribution does not differ significantly from a uniform distribution at two or more frequencies are not listed. All angle differences are given in degrees.}
 \label{tab:evpas_1-3-opt}
 \begin{tabular}{c|c|c|c|c|c|c|c}
    \hline
    B1950 & J2000 & 3~mm-1~mm & 3~mm-Opt & 1~mm-Opt & 3~mm-PA & 1~mm-PA & Opt-PA \\
   & &    [deg] & [deg] & [deg] & [deg] & [deg] & [deg]\\
    \hline
0219+428   & J0222+4302   &    -- &   0.0 $\pm$   2.7 & -- &   4.2 $\pm$   6.1 & -- &   4.2 $\pm$   5.5 \\
0336$-$019 & J0339$-$0146 &    -- &   7.5 $\pm$   4.6 & -- &   3.3 $\pm$   6.2 & -- &   4.2 $\pm$   5.0 \\
0420$-$014 & J0423$-$0120 & -- &-- &  27.3 $\pm$   5.7 & -- &  44.3 $\pm$   6.4 &  71.6 $\pm$   4.3  \\
0735+178   & J0738+1742   &    -- &  33.7 $\pm$   5.4 & -- &  67.1 $\pm$  11.7 & -- &  79.2 $\pm$  11.0 \\
0829+046   & J0831+0429   &  14.6 $\pm$   5.3 &  16.8 $\pm$   3.0 &   2.2 $\pm$   4.7 &   1.7 $\pm$   6.3 &  12.9 $\pm$   7.3 &  15.1 $\pm$   5.9 \\
0851+202   & J0854+2006   &   1.0 $\pm$   2.4 &   5.5 $\pm$   1.5 &   4.5 $\pm$   2.1 &  10.3 $\pm$   1.8 &  11.3 $\pm$   2.3 &  15.8 $\pm$   1.3 \\
0954+658   & J0958+6533   &   1.0 $\pm$   2.9 &   8.3 $\pm$   2.1 &   9.3 $\pm$   2.5 &  13.8 $\pm$   2.8 &  14.8 $\pm$   3.1 &   5.5 $\pm$   2.3 \\
1055+018   & J1058+0133   &   6.8 $\pm$   4.7 &  74.8 $\pm$   4.3 &  67.9 $\pm$   4.5 &  19.0 $\pm$   7.2 &  25.8 $\pm$   7.4 &  86.2 $\pm$   7.1 \\
1101+384   & J1104+3812   &    -- &  24.8 $\pm$   4.8 & -- &  15.2 $\pm$  15.0 & -- &   9.6 $\pm$  14.2 \\
1219+285   & J1221+2813   &    -- &   5.3 $\pm$   4.2 & -- &  31.8 $\pm$  11.4 & -- &  37.1 $\pm$  10.6 \\
1222+216   & J1224+2122   &   1.6 $\pm$   3.2 &   1.8 $\pm$   2.0 &   3.4 $\pm$   2.7 &  25.5 $\pm$   2.7 &  27.0 $\pm$   3.2 &  23.7 $\pm$   2.0 \\
1226+023   & J1229+0203   &   7.1 $\pm$   2.8 &  89.9 $\pm$   1.6 &  82.8 $\pm$   2.9 &  80.5 $\pm$   1.2 &  87.6 $\pm$   2.7 &   9.7 $\pm$   1.4 \\
1253$-$055 & J1256$-$0547 &   6.0 $\pm$   4.1 &  18.4 $\pm$   2.6 &  12.4 $\pm$   3.3 &   7.4 $\pm$   2.7 &  13.5 $\pm$   3.4 &  25.8 $\pm$   1.2 \\
1308+326   & J1310+3220   &   8.1 $\pm$   6.5 &  23.7 $\pm$   4.8 &  15.6 $\pm$   6.0 &   6.4 $\pm$   6.1 &   1.7 $\pm$   7.1 &  17.3 $\pm$   5.5 \\
1406$-$076 & J1408$-$0752 &    -- &  69.0 $\pm$   6.5 & -- &  23.8 $\pm$   8.5 & -- &  87.2 $\pm$   8.5 \\
1611+343   & J1613+3412   &    -- &  81.3 $\pm$   4.4 & -- &  89.0 $\pm$   5.0 & -- &   9.6 $\pm$   4.0 \\
1641+399   & J1642+3948   &    -- &  14.6 $\pm$   2.8 & -- &   5.2 $\pm$   3.6 & -- &  19.8 $\pm$   2.9 \\
1730$-$130 & J1733$-$1304 &    -- &  21.2 $\pm$   5.5 & -- &  36.0 $\pm$   4.5 & -- &  57.2 $\pm$   3.8 \\
1749+096   & J1751+0939   &    -- &   2.0 $\pm$   4.2 & -- &  22.6 $\pm$   4.5 & -- &  20.7 $\pm$   2.8 \\
2200+420   & J2202+4216   &   5.1 $\pm$   1.5 &   0.3 $\pm$   1.0 &   5.4 $\pm$   1.3 &   5.6 $\pm$   2.0 &   0.4 $\pm$   2.2 &   5.9 $\pm$   1.9 \\
2223$-$052 & J2225$-$0457 &   3.5 $\pm$   4.0 &  68.5 $\pm$   4.1 &  72.0 $\pm$   4.7 &  29.8 $\pm$   3.9 &  33.3 $\pm$   4.6 &  38.7 $\pm$   4.7 \\
2230+114   & J2232+1143   &   9.9 $\pm$   4.0 &  67.2 $\pm$   2.5 &  77.2 $\pm$   3.6 &  63.0 $\pm$   4.2 &  53.1 $\pm$   5.0 &  49.8 $\pm$   3.8 \\
2251+158   & J2253+1608   &   0.8 $\pm$   4.1 &  47.2 $\pm$   3.0 &  46.4 $\pm$   3.1 &   2.3 $\pm$   3.3 &   1.5 $\pm$   3.4 &  44.9 $\pm$   1.9 \\
    \hline
  \end{tabular}
 \end{table*}

%%%%%%%%%%%%%%%

The list of sources with EVPAs distributed non-uniformly in at least two observing bands is reported in Tab.~\ref{tab:evpas_1-3-opt}. In these sources, we can investigate the alignments of mean EVPA orientations across bands. We notice a general agreement between 1 and 3~mm EVPA orientations. In the majority of BL Lacs (64\%), mm-EVPAs are also closely aligned (within 10$^{\circ}$, considering 1-$\sigma$ uncertainty) with optical EVPAs, while this happens in the minority (19$\%$) of FSRQs. The alignment of optical or mm-EVPAs with the jet PA is observed in almost half of the BL Lac objects, and a small percentage of FSRQs, 14 and 24$\%$ respectively for the optical and mm-case.

In two FSRQs, J1229+0203 and J1613+3412, optical and mm-EVPAs are almost perpendicular to each other. Interestingly, in both cases, the mm-EVPAs are also perpendicular to the jet PA while the optical EVPAs are within 10$^{\circ}$ from the jet PA.

We computed $p_0$, as described in Sec.~\ref{sec:data}, for all sources and bands, and we found that in general the fractional polarization correlates with the observing frequency, with $p_{0}^{3~mm} < p_{0}^{1~mm} < p_{0}^{optical}$. Another result is that BL Lac objects have higher average fractional polarization than FSRQs at all bands; the values of $p_0$ and relative errors are reported in Tab~\ref{tab:mean_p}.

% \cite{Marscher&Jorstad}, found that in six of the blazars in their sample (0219+428, 0851+202, 0954+658, 1253-055, 1652+398, and 2200+420), optical EVPAs were often stable for years at a preferential orientation within $\sim\pm$20$^{\circ}$ from the jet position angle. 

\begin{table}[]
    \centering
    \caption{The average intrinsic mean fractional polarization and its standard error for FSRQs and BL Lac objects at the three frequencies.}
    \begin{tabular}{c|c|c|c}
     \hline
     Class & $p_{0}^{3~mm}$ & $p_{0}^{1~mm}$ & $p_{0}^{opt}$\\
     & [\%] & [\%] & [\%]\\
     \hline
     BL Lacs    &  4.3 $\pm$ 0.5 & 5.9 $\pm$ 0.5 & 9.8 $\pm$ 0.8\\
     FSRQs      &  3.3 $\pm$ 0.2 & 4.7 $\pm$ 0.3 & 7.3 $\pm$ 1.0\\
    \end{tabular}
    \label{tab:mean_p}
\end{table}

%%%%%%%%%%%%%%%%%%%%%%%%%%%%%%%%%%%%%%%%%%%%%%%%%%

\subsection{EVPA Variability}
To measure the spread of EVPAs across the different observing bands and object classes, we computed the circular standard deviation of EVPAs per source and frequency. The mean standard deviation values and associated standard errors for the FSRQs and BL LAC object class are reported in Table~\ref{tab:mean_EVPAstd}. The mean standard deviation values are higher in FSRQs than in BL Lacs and increase with the observing frequency. However, it is worth noting that all measurements are consistent with each other within 1-$\sigma$, except for the mean standard deviation of optical EVPAs in  FSRQs.

\begin{figure}
    \includegraphics[width=0.99\hsize]{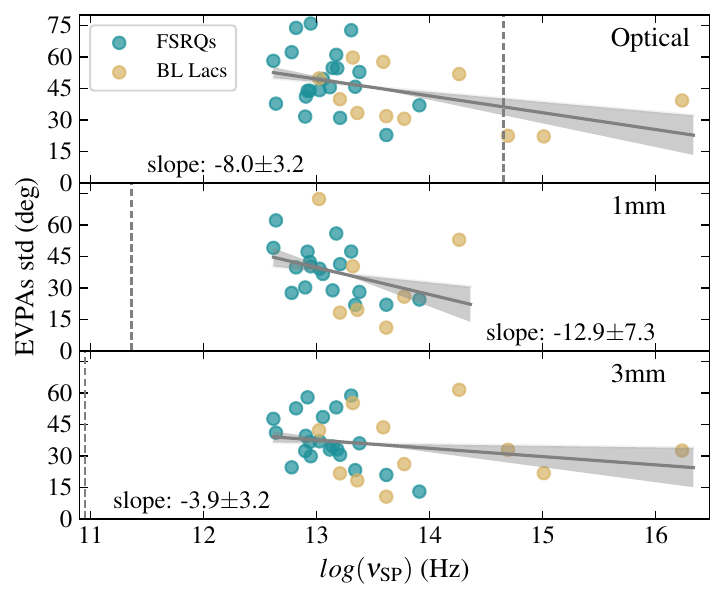}
\caption{Dependence of the EVPA spread on the synchrotron peak position. The vertical lines in each panel indicate the respective observing frequencies. NOTES: The trend observed at 1 and 3~mm, resembling the trend at optical wavelength, is similar to the trend in \cite{Angelakis2016} and fits with the model proposed in \cite{Potter2015}. The gray line and the shadow area represent the best linear fit using a least-squares fitting method and the 1-$\sigma$ deviation, respectively.}
\label{fig:EVPA_SPP}
% polami/12_plot_opt_mm_pol/plot_EVPA_degP_SPP.py
\end{figure}

\begin{figure}
% produced by plot_pol_var_vs_radio_lumin.py
% in ./18_jet_power_pol_var/
    \includegraphics[width=0.99\hsize]{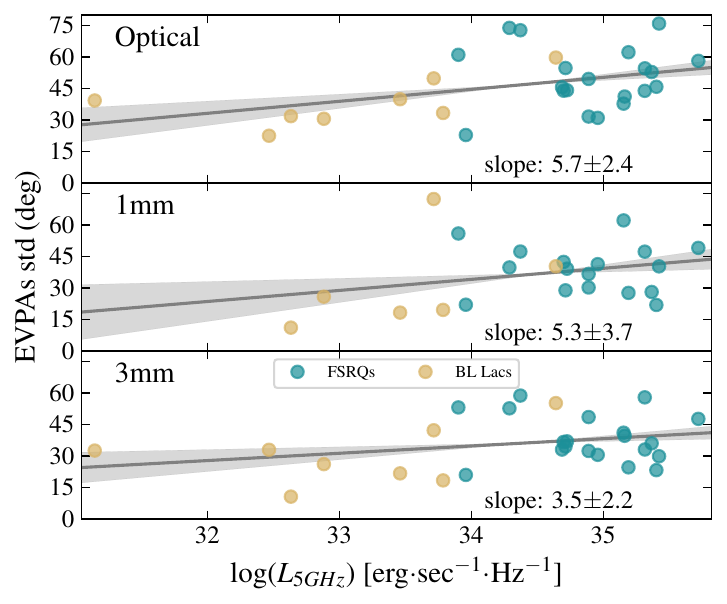}
\caption{EVPAs spread in three bands as a function of radio luminosity. The gray line and the shadow area represent the best linear fit using a least-squares fitting method and the 1-$\sigma$ deviation, respectively.}
\label{fig:EVPAstd_vs_RadioLumin}
\end{figure}

\begin{figure}
% produced by plot_pol_var_vs_gamma_lumin.py
% in ./13_gamma_lum_mm_pol_var/
    \includegraphics[width=0.99\hsize]{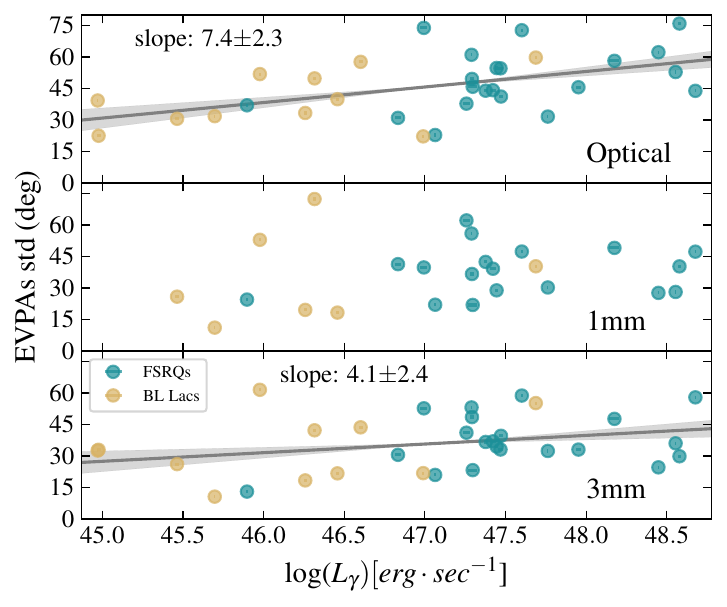}
\caption{EVPAs spread in three bands as a function of gamma-ray luminosity. The gray line and shadow area are as in Fig.~\ref{fig:EVPAstd_vs_RadioLumin}.}
\label{fig:EVPAstd_vs_GammaLumin}
\end{figure}

We investigated the dependence of the EVPA standard deviation on the synchrotron peak frequency, $\nu_{\rm SP}$ (Fig.~\ref{fig:EVPA_SPP}), as well as on the radio (Fig.~\ref{fig:EVPAstd_vs_RadioLumin}), and gamma-ray (Fig.~\ref{fig:EVPAstd_vs_GammaLumin}) luminosities. Here $\nu_{SP}$ refers to the synchrotron peak frequency in the $\nu$$F_\nu$ SED. Radio and gamma-ray luminosities are taken as proxies of the jet power. The radio monochromatic luminosity at 5 GHz is taken from \cite{Kudryavtsev2024}, while the $\gamma$-ray luminosity was computed using equation 3 of \cite{Singal2014}, with energy flux measurements obtained from the Fermi Large Area Telescope Fourth Source Catalog Data Release 4 (4FGL-DR4) \citep{Abdollahi2022}. Synchrotron peak frequency values in the observer frame are taken from \cite{Ajello2020}, and redshift measurements are from \cite{Agudo2018I}. Redshift of J0721+7120 is taken from \cite{Pichel2023}.

We find an anti-correlation between the EVPA spread and $\nu_{\rm SP}$ across all three observing bands, with the strongest correlation observed in the optical data\footnote{Hereafter, we consider a correlation significant when the slope of the best-fit line deviates from zero by more than its 1$\sigma$ uncertainty.}. %(2-$\sigma$ deviation from zero instead of 1-$\sigma$). 
Given the observed anti-correlations between the synchrotron peak and both jet power and Compton dominance in blazars \citep{Ghisellini1998}, we would expect a positive correlation between the EVPA spread and both the radio and gamma-ray luminosities. Such a correlation is indeed found in data: Figs.~\ref{fig:EVPAstd_vs_RadioLumin} and ~\ref{fig:EVPAstd_vs_GammaLumin} show the positive correlations between the EVPA standard deviations and the radio and gamma-ray luminosities, which are again stronger in optical data. Only at 1~mm the best-fit between the EVPA spread and the gamma-ray luminosity is consistent with no correlation within 1-$\sigma$. 
%We also tested the correlation between the EVPA spread and the jet power, the latter computed following the prescription in \cite{Foschini2024}. In that case, we found a positive correlation only for optical data, which is also weaker than the correlation found using radio and gamma-ray luminosities (see Appendix~\ref{app:B}). 

\begin{figure}
% produced by plot_evpa_diff.py
% in ./12_plot_opt_mm_pol/
    \includegraphics[width=0.99\hsize]{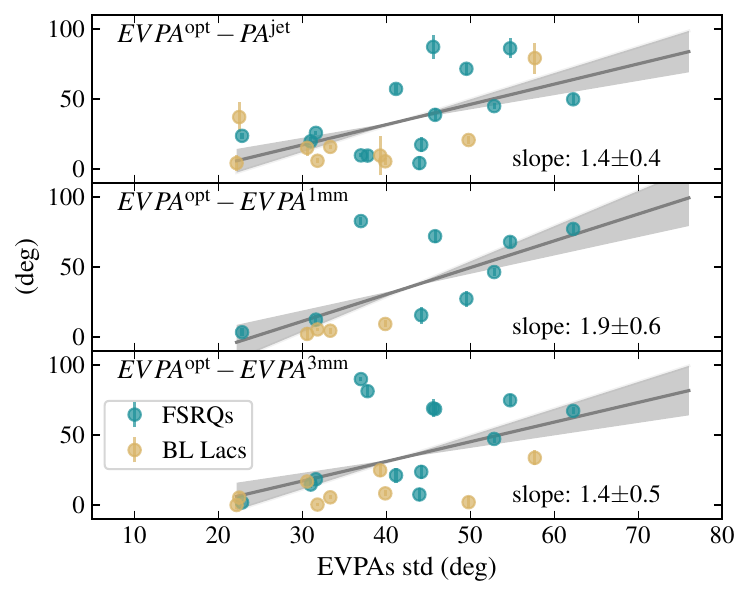}
\caption{The optical EVPA spread in function of the alignment between the optical EVPA and the jet PA \textit{(upper panel)}, the 1~mm EVPAs \textit{(middle panel)}, and the 3~mm EVPAs \textit{(bottom panel)}. The optical EVPA spread correlates with the divergence between optical EVPAs and the three quantities explored in the three panels. The gray line and the shadow area represent the best linear fit using a least-squares fitting method and the 1-$\sigma$ deviation, respectively.}
\label{fig:Opt_evpa}
\end{figure}

From the analysis of EVPA orientations across bands (see sect.~\ref{sec:evpa_pa}), we see that optical EVPAs in FSRQs tend to be less aligned with jet PAs and also with mm-EVPAs than in BL Lacs. We investigated the possible role of the EVPA variability in leading the observed differences. We found a positive correlation between the optical EVPA divergences and the optical EVPA spread, as shown in Fig.~\ref{fig:Opt_evpa}. We see that sources with more stable optical EVPAs tend to have optical EVPAs more aligned with both the jet PA and mm-EVPAs. The majority of BL Lacs and FSRQs are respectively located at the left and right ends of the plot in Fig.~\ref{fig:Opt_evpa}.

\begin{table}[]
    \centering
    \caption{The mean EVPA circular standard deviations and associated standard errors computed at the three frequencies for FSRQs and BL Lac objects.}
    \begin{tabular}{c|c|c|c}
     \hline
     Class & std-3~mm & std-1~mm & std-Opt\\
     & [deg] & [deg] & [deg]\\
     \hline
     BL Lacs    &  33.0 $\pm$ 4.6 & 32.98 $\pm$ 6.3 & 39.8 $\pm$ 3.9\\
     FSRQs      & 37.3 $\pm$ 2.6 & 37.9 $\pm$ 2.5 & 49.4 $\pm$ 3.1\\
    \end{tabular}
    \label{tab:mean_EVPAstd}
\end{table}

%%%%%%%%%%%%%%%%%%%%%%%%%%%%%%%%%%%%%%%%%%%%%%%%%%
\subsection{Flux Density and Fractional Polarization Variability}
\label{modindex}
We parametrized the variability of the total flux density and fractional polarization using the intrinsic modulation indices $m_{\rm fl}$ and $m_{\rm pol}$, calculated as described in Sec.~\ref{sec:data}. We examined their relationships with the EVPA distribution spread and the intrinsic average fractional polarization $p_0$.

\begin{figure}
% produced by plot_EVPAstd_vs_flux_mod_ind.py
% in ./17_evpa_vs_mod_ind_pol/
    \includegraphics[width=0.99\hsize]{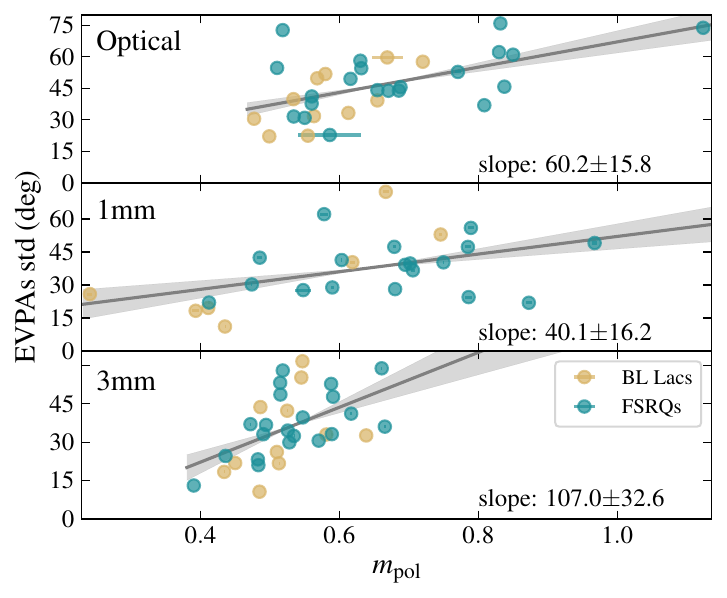}
\caption{Intrinsic modulation index of fractional polarization vs the standard deviation of EVPA in the corresponding band.}
\label{fig:EVPAstd_vs_PolModInd}
% polami/17_evpa_vs_mod_ind_pol/plot_mod_ind_vs_evpa_var.py
\end{figure}

No correlation was found between EVPA spread and $m_{\rm fl}$, while $m_{\rm pol}$ positively correlated with the EVPA spread, as shown in Fig.~\ref{fig:EVPAstd_vs_PolModInd}: the greater the variability in fractional polarization, the broader the range of EVPAs observed.

\begin{figure}
    \includegraphics[width=0.99\hsize]{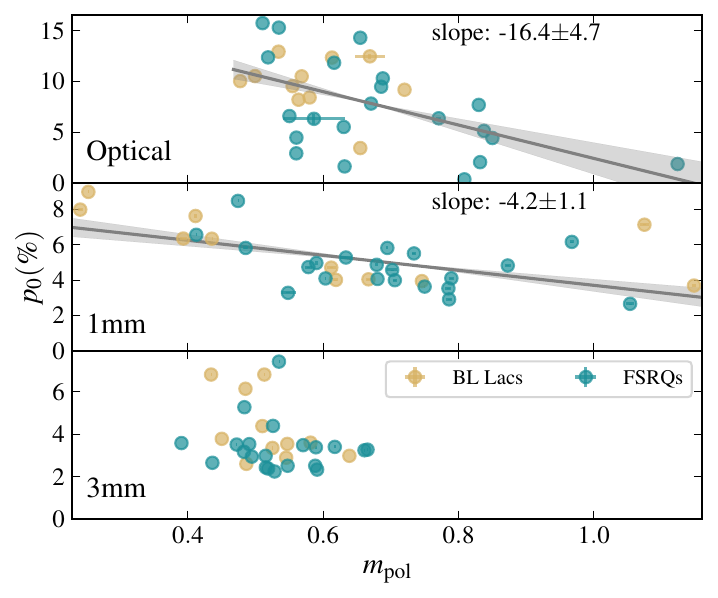}
\caption{Intrinsic average fractional polarization vs the intrinsic modulation index of polarization in the same band.}
\label{fig:DegPD_vs_PolModInd}
% polami/16_mod_ind_vs_pol/plot_mod_ind_vs_av_pol.py
\end{figure}

\begin{figure}
    \includegraphics[width=0.99\hsize]{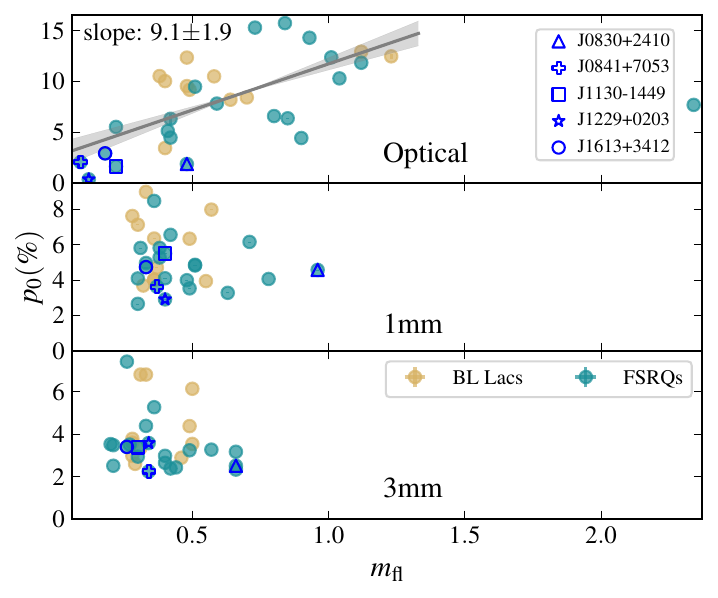}
\caption{Intrinsic modulation index of the flux density vs the intrinsic average polarization in the corresponding band. The outlier (CTA102 a.k.a. J2232+1143) in the optical panel is not taken into account in the fitting procedure.}
\label{fig:DegPD_vs_FluxModInd}
% polami/15_flux_var_vs_pol_var/plot_av_pol_vs_flux_mod_ind.py
\end{figure}

The intrinsic average fractional polarization $p_0$ was also analyzed in function of $m_{\rm pol}$ (Fig.~\ref{fig:DegPD_vs_PolModInd}) and $m_{\rm fl}$ (Fig.~\ref{fig:DegPD_vs_FluxModInd}). The outlier in Fig.~\ref{fig:DegPD_vs_FluxModInd} is J2233+1143 (CTA~102), which was excluded from the data fitting. We believe the high $m_{\rm fl}$ in CTA~102 is dominated by the exceptional optical flare it underwent at the end of 2016, during which its brightness increased by 6 magnitudes, making it the most luminous blazar ever observed in the optical sky \citep{Raiteri2017Natur}.

At optical and 1~mm wavelengths, $p_0$ showed a negative correlation with $m_{\rm pol}$, and a positive correlation with $m_{\rm fl}$ in the optical data only. No correlation was found between $p_0$ and any of the modulation indices at 3~mm. The tendency for sources to exhibit lower $p_0$ and lower $m_{\rm fl}$ at optical wavelengths may suggest that a turbulent magnetic field dominates at those wavebands. However, a pure turbulence scenario could not account for the $p_0$-$m_{\rm pol}$ anti-correlation, which instead predicts that a small number of cells leads to both high $p_0$ and $m_{\rm pol}$ \citep{Marscher2014}. Another possible explanation could be a different origin for the polarized emission at these frequencies. For example, in some sources, the optical polarized flux may have a significant contribution from non-beamed or thermal components, leading to a decrease in fractional polarization below a few percent and resulting in lower flux density variability. 

The positive correlation found in optical between $p_0$ and  $m_{\rm fl}$ suggests the tendency for sources to show high degree of polarization during prominent optical flares. Such a behavior has been occasionally observed in several blazars during single flare events \citep[e.g.,][]{Marscher2010, Casadio2015}.  

The lack of correlation between the polarization degree and variability at 3~mm suggests a magnetic field structure that is less subject to extreme variability. This is confirmed by the 3~mm $m_{\rm pol}$ and $m_{\rm fl}$ values, which are generally lower ($<$0.7) than the values at 1~mm and optical bands. This is in agreement with the POLAMI results presented in \cite{Agudo2018III}, where 1~mm data were found to be more variable than 3~mm data both in polarization and total intensity.

%\section{Luminosity vs polarization variability} \label{sec:lum_pol}
%\subsection{Radio luminosity vs polarization variability} \label{sec:lum_rad_pol}
%In Figures~\ref{fig:DegPD_vs_RadioLumin} and \ref{fig:EVPAstd_vs_RadioLumin} we demonstrate dependence of the median fractional polarization and EVPA standard deviation in different bands as a function of source's radio luminosity.
%\begin{figure}
%    \includegraphics[width=0.99\hsize]{./figures/DegPD_vs_RadioLumin.pdf}
%\caption{Average fractional polarization in three bands as a function of radio luminosity.}
%\label{fig:DegPD_vs_RadioLumin}
%\end{figure}

\section{Discussion} \label{sec:disc}

\subsection{On the Origin of Optical and mm-Wave Emission in Blazars.}
The comprehensive analysis of optical and millimeter-wave polarimetric data for the 21 FRSQs and 11 BL Lacs monitored under the POLAMI program has revealed class- and frequency-dependent correlations. Some of the reported correlations were also previously found by \cite{Jorstad2007}, who conducted a similar optical-millimeter comparative study on a subsample of the sources considered in this work. Despite the smaller blazar sample (15 sources) and the shorter data coverage ($\leq$4 years) analyzed in \cite{Jorstad2007}, the authors still found that BL Lacs tend to exhibit better alignment of polarization position angles between frequencies and with the jet position angle than FSRQs. It is important to note that the results in \cite{Jorstad2007} are based on single-epoch comparisons, rather than on integrated quantities as in the present study. The fact that blazars exhibit similar behavior on different timescales strengthens the conclusions of the two optical-millimeter comparative studies and highlights the intrinsic differences in magnetic field configuration and/or emission-site location between the two blazar classes. It is therefore clear that an appropriate physical model of blazars should be able to account for these differences.

Recent results from the Imaging X-ray Polarimetry Explorer \citep{Marscher2024} suggest the "energy-stratified" model as the best framework for explaining the polarimetric behavior of blazars across the optical to X-ray bands. In this model, relativistic particles in jets are accelerated in the proximity of the front of shock waves that are either stationary (recollimation shocks) or travel along the jet \citep{Sciaccaluga2025}. As shock-accelerated particles propagate downstream, they undergo progressive cooling, primarily via synchrotron and Compton radiation. This leads to the energy stratification of electrons as they travel away from the acceleration site. In addition, shock compression is expected to amplify the magnetic field component parallel to the shock front, and therefore to have a magnetic field that is partially ordered. One thus expects a higher degree of polarization at high frequency, close to the shock front, and lower at low frequencies, further down where the magnetic field becomes more turbulent. 

This model may account for the frequency dependence of the degree of polarization, as well as the negative correlation found between EVPA spread and $\nu_{SP}$ in the optical domain (Fig.~\ref{fig:EVPA_SPP}). The same correlation was in fact found in a different sample of blazars using optical polarimetric data from the Robopol program \citep{Angelakis2016}, and it was explained by accounting for distinct locations for the optical emission in low (LSP, log$(\nu_{SP})<$14), and high-synchrotron-peaked (HSP, log$(\nu_{SP})>$15) blazars. In LSP sources, such as FSRQs and low-frequency peaked BL Lac objects, the optical emission probes the high-energy end of the synchrotron spectrum, which typically peaks in the near-infrared. According to the energy-stratified model, the optical emission in these sources originates from a compact region close to the shock front, where higher and faster variability is expected both in total and linearly polarized emission. In contrast, in HSP blazars the optical emission corresponds to the low-energy end of the synchrotron component and thus arises further downstream from the particle-accelerated site, where the magnetic field is more stable. 
%In this scenario, the optical EVPA variability in LSP is primarily driven by the passage of plasmoids --- regions of enhanced magnetic field or electron density --- across the shock front, whereas in HSP it is mostly due to stochastic variations. 
However, the same qualitative explanation is hard to reconcile with the EVPA spread - $\nu_{SP}$ correlation found also at millimeter waves, since mm-wave emission comes from the low-energy part of the synchrotron spectrum in both spectral classes.

The correlations between EVPA spread and $\nu_{SP}$, at both optical and millimeter-wave, can be interpreted in an alternative energy stratified model, where the stratification results from magnetic reconnection in a turbulent environment \citep{deJonge2025, Zhang2025}. Electrons close to and above $\nu_{SP}$ concentrate in small parts of plasmoids or plasmoid mergers, where the magnetic field can be very ordered, leading to high $p_0$ and low $m_{pol}$. Such small regions evolve rapidly: electrons can quickly move to other parts of plasmoids and mergers, where the field lines are ordered in a different direction \citep{Zhang2018}. This can result in large EVPA spreads or even angle rotations. By contrast, electrons below $\nu_{SP}$ occupy most of the plasmoid volume and represent its homogeneous component. This results in low $p_0$ and small EVPA spread \citep{Zhang2024}. In summary, for a given synchrotron peak frequency, the average $p_0$ and EVPA spread rise with the electron energy. The above scenario naturally explains the anti-correlations between EVPA spread and $\nu_{SP}$ as well as between $p_0$ and $m_{pol}$ at all wavelengths. However, this model predicts an anti-correlation between $p_0$ and $\nu_{SP}$, which is not observed in this study, and it does not reproduce the higher $p_0$ values of BL Lacs compared to FSRQs. 

The multi-band EVPA spread - $\nu_{SP}$ anti-correlation may also indicate a more ordered global magnetic field geometry in low-power sources compared to high-power ones.  Given the known anti-correlation between jet power and $\nu_{SP}$ in blazars, the EVPA spread - $\nu_{SP}$ anti-correlation translates into a positive correlation of EVPA spread with the jet power, here represented by the radio and gamma-ray luminosities (Figs.~\ref{fig:EVPAstd_vs_RadioLumin} and ~\ref{fig:EVPAstd_vs_GammaLumin}). In the jet emission model proposed by \cite{Potter2015}, high-power blazars (FSRQs) exhibit jets that accelerate over longer distances, achieving higher asymptotic Lorentz factors and larger transition radii. The transition radius marks the point where the jet reaches equipartition and transits from an accelerated to a freely expanding geometry. The bulk of the synchrotron emission is expected to originate at the transition region, where the jet attains its maximum Lorentz factor. Given the larger transition radii, in FSRQs the transition region spans a larger area, resulting in a lower magnetic field strength and, consequently, a lower $\nu_{SP}$. This model would then explain the anti-correlation between jet power and $\nu_{SP}$ found in blazars. 
Additionally, larger transition radii encompass a broader range of magnetic field orientations, which may account for the lower fractional polarization and larger EVPA spread observed in FSRQs in this study. However, this model fails to provide a satisfactory explanation for the high $m_{pol}$ observed in these sources.
 
%The correlations found between optical quantities are in good agreement with this model. In fact, we found that larger optical EVPA spreads are associated with poorer alignments between EVPAs across bands, higher polarization variability, and a lower degree of polarization, consistent with the optical emission in high-power sources arising from a larger volume than in low-power ones.

We have examined the behaviour of sources with very low optical fractional polarization, p$_0\leq3\%$, which are also marked in Fig.~\ref{fig:DegPD_vs_FluxModInd}. Differently, at millimeter wavelengths, these sources have p$_0\geq2\%$. As explained in Sec.~\ref{modindex}, the unusually low degree of polarization in the optical band can either hint at a more turbulent magnetic field in the optical emitting region or at a significant contribution to the optical emission from the non-beamed or thermal components, like the accretion disk, BLR, or torus. 
%From Figs.~\ref{fig:DegPD_vs_FluxModInd} and ~\ref{fig:DegPD_vs_PolModInd} we learn that sources with low optical p$_0$ in optical tend to have higher $m_{\rm pol}$ and lower $m_{\rm fl}$ in optical than sources with high p$_0$.
From Figs.~\ref{fig:DegPD_vs_FluxModInd} we learn that sources with low optical p$_0$ tend to vary less in flux density, which could be explained if the optical emission comes from a non-beamed component. 
The optical $m_{\rm pol}$ and EVPA spread instead cover a wide range of values, from 0.6 to 1.1 for $m_{\rm pol}$, and from 37$^{\circ}$ to 76$^{\circ}$ for the EVPA spread, or even EVPAs uniformly distributed as is the case of J1130-1449. 
To investigate the possible contribution of the disk, or BLR, typically located upstream of the mm-wave emitting region, we inspected the very-long-baseline-interferometry (VLBI)-\textit{Gaia} offsets for the POLAMI sources, taking values from \cite{Blinov2024}. The milliarcsecond (mas) precision in optical astrometry provided by the European Space Agency’s (ESA) Gaia mission \citep{Gaia2016A&A} led to the discovery of AGNs exhibiting significant positional offsets between radio VLBI and Gaia measurements, with amplitudes reaching tens of milliarcseconds \citep[e.g.][]{Petrov2017MNRAS}.
It was found that in jetted AGN, the VLBI-to-Gaia displacements (VGDs) tend to align with the direction of the radio jet. Moreover, jetted-AGN with \textit{Gaia} positions located upstream of the VLBI coordinates, hence toward the disk/BLR region, exhibit lower optical polarization than their downstream counterparts, as expected if disk or BLR dominate the optical emission \citep{Kovalev2020, Blinov2024}. 

\begin{figure}
\center
% produced by DP_vs_VG.py
% in ./9_VG_offsets/
    \includegraphics[width=0.99\columnwidth]{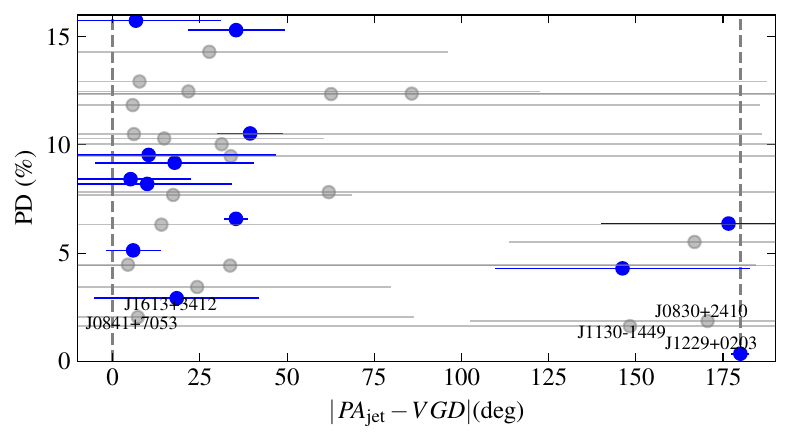}
\caption{Dependence of optical fractional polarization on the angle between the jet position angle and the VLBI-to-Gaia displacement vector.}
\label{fig:VG}
\end{figure}
In \cite{Blinov2024}, the VGD is parametrized by the angle between the VLBI jet direction and the vector connecting the
VLBI and Gaia positions. In this way, the 0$^{\circ}$ and 180$^{\circ}$ angles represent sources with \textit{Gaia} positions downstream and upstream of the VLBI coordinates, respectively. We adopted the definition of the uncertainty on the VGD angle from \cite{Blinov2024}, but reduced its value by half, as the original values refer to 2-$\sigma$ positional errors and may therefore be overestimated. The VGDs for the POLAMI sources are shown against their mean optical fractional polarization, in Fig.~\ref{fig:VG}. Sources with VGD uncertainties $>$~90$^{\circ}$ are marked in gray. We noted a bi-modal distribution in VGD, around 0$^{\circ}$ and 180$^{\circ}$, in agreement with previous studies  \citep{Kovalev2020, Blinov2024}. Interestingly, of the five sources with p$_0\leq3\%$ (see Fig.~\ref{fig:DegPD_vs_FluxModInd}), three have VGDs close to 180$^{\circ}$. There is also a clear lack of sources with high polarization and upstream offsets. All this may support the hypothesis of a strong contribution from the accretion disk or BLR to both the total and linearly polarized emission at optical frequencies. 

\cite{Williamson2014} reached similar conclusions from the analysis of optical spectral indices over different periods of activity of the sources. In particular, they found a flattening of the optical spectra of FSRQs during quiescent states, but not in BL Lacs. This suggests a dominant disk contribution in optical emission in FSRQs during quiescent states. Notably, the three sources mentioned above, which display low p$_0$ and VGDs close to 180$^{\circ}$, also show low $m_{\rm fl}$, suggesting a dominance of quiescent states.  
However, the contribution from the accretion disk is well documented in only one of the three sources, J1229+0203. According to \cite{Li2020}, the jet accounts for only 10-40\% of the optical emission in this source, depending on the brightness state.

\subsection{The Magnetic Field Structure.}

Of the six sources with 3mm-EVPAs stable around the mean for at least 70\%  of their time, three (J1229+0203, J1256-0547, J2202+4216) show either a zig-zag pattern in their jet profiles \citep{Kravchenko2025} or a filamentary structure in both total and polarized emission, with consequent variations of EVPA orientations along the filaments~\citep{Lobanov2001Sci...294..128L, Fuentes2023, Kim2023}. The stability of 3~mm~EVPAs suggests that integrated values of the fractional polarization, as measured by the POLAMI program, are markers of the bulk magnetic field and are not affected by local perturbations.

Several sources have shown evidence of toroidal or helical magnetic field geometries in previous studies: J0854+2006 \citep{Gomez2022}, J1229+0203 \citep{Hovatta2012}, J1256-0547 \citep{Fuentes2023}, J2202+4216 \citep{Gomez2016}, J2232+1143 \citep{Casadio2019}.
Of the above sources, J0854+2006, J1256-0547, and J2202+4216 show alignment of EVPAs among frequencies and with the jet PA, in agreement with the expected helical magnetic field geometry. In the two remaining sources, J1229+0203 and J2232+1143, mm-EVPAs are instead almost perpendicular to the jet PA and deviate from optical EVPA directions. \cite{Hovatta2019} found a very high Faraday rotation measure (RM) of $(5.0 \pm 0.3) \times 10^5$ rad m$^{-2}$ at 1~mm in the jet of J1229+0203, leading to an EVPA rotation of $\sim$30$^{\circ}$. Even considering the large rotation affecting 1~mm EVPAs, this could not account for the almost 90$^{\circ}$ difference between millimeter and optical EVPAs or millimeter EVPAs and jet PA. This implies that millimeter EVPAs as measured within POLAMI do not trace the overall helical magnetic field geometry, in this source. The polarized emission at mm-wave could be dominated by a bright feature observed downstream of the 3~mm core, as observed in \cite{Casadio2017}. 
The discrepancy between optical and mm-EVPAs is additional evidence supporting a disk/BLR contribution to the optical emission.

\section{Summary and Conclusions} \label{sec:sum}

We present a systematic study of a blazar sample, comparing optical and millimeter-wavelength polarimetric data from 15 years of observations. This work was carried out within the framework of the POLAMI program, which defined the source sample (21 FSRQs, 11 BLLacs, and 2 Radio galaxies) and provided polarimetric data at 1 and 3~mm from October 2006 to October 2021. These data were complemented with optical photo-polarimetric data obtained from several telescopes (see \S~\ref{sec:data}), covering the period from April 2005 to September 2021. The main results can be summarized as follows:
\begin{itemize}
\item The average fractional polarization increases with observing frequency;

\item On average, BL Lacs have higher fractional polarization than FSRQs;

%\textcolor{blue}{Better-ordered magnetic field at higher frequencies? Either I) the polarized flux is not co-spatial at the three frequencies, or II) the higher the frequency, the more compact the emitting region. In the latter, we may expect an increase in fractional polarization with frequency if multiple regions with different magnetic field orientations are present, like in the TEMZ model. However, the EVPA orientations at the three frequencies may still be close to each other if, above all variable local magnetic fields, there is a dominant feature representing the bulk magnetic field orientation in that region. If the emission at the different frequencies is not co-spatial (case I), we wouldn't necessarily expect similar EVPA orientations at the three frequencies.}

\item 67\% of FSRQs and 82\% of BL Lacs, have EVPAs following a non-uniform distribution. The EVPAs in both radio galaxies instead follow a uniform distribution;

\item EVPAs in BL Lacs are less variable (i.e., have a smaller spread) than in FSRQs at all three bands. In the majority of BL Lacs, EVPAs tend to align across bands, whereas in FSRQs, this occurs in the minority of sources;

\item The larger the spread in optical EVPAs, the greater the deviation between optical and mm-wave EVPAs, as well as between optical EVPA and the jet PA;

\item The EVPA spread at the three frequencies anti-correlates with the synchrotron peak frequency;

\item Of the five sources with a low degree of polarization ($<$3\%) and low flux density variability in optical, three of them have \textit{Gaia} positions upstream of the VLBI radio core;

\end{itemize}

The above findings are in good agreement with the model proposed by \cite{Potter2015}, where the contrasting observational properties of FSRQs and BLLacs are attributed to differences in accretion rate, and consequently in jet power. High-power sources can propagate to larger distances before reaching equipartition, resulting in larger jet radii at the transition regions compared to low-power sources. In this framework, and considering the above findings, the millimeter and optical emissions arise from a larger jet section in FSRQs than in BLLacs, with consequently lower magnetic field strength that would lead to a lower synchrotron peak frequency, lower degree of polarization, and larger spread in EVPA orientation, as we observe in this study. 

The observed positive correlation of polarization degree with frequency indicates an increasing compactness of the emitting region with frequency. This, and some of the observed correlations in polarization, may also be explained within an energy stratified model in a turbulent environment, where the stratification arises from magnetic reconnection \citep[e.g.][]{deJonge2025}. Alternatively, similar behavior could result from turbulence \citep{Marscher2014} in an energy-stratified shock-in-jet model \citep{Liodakis2022}. It is likely that a hybrid approach, incorporating models that address both macro- and micro-scale processes, offers the most realistic representation. 

Additionally, we find that some of the sources with low optical fractional polarization (all FSRQs) show strong evidence that the optical and mm-wave emission are not co-spatial, with the optical emission likely having a significant contribution from the accretion disk and/or BLR.

%% Please use the acknowledgment and contribution environments. This will 
%% be anonomyized when the "anonymous" style option is used. 
\begin{acknowledgments}
The POLAMI observations were carried out at the IRAM 30m Telescope. IRAM is supported by INSU/CNRS (France), MPG (Germany), and IGN (Spain). Some of the optical data analyzed in these paper were collected at the Centro Astron\'{o}mico Hispano en Andaluc\'ia (CAHA); which is operated jointly by Junta de Andaluc\'{i}a and Consejo Superior de Investigaciones Cient\'{i}ficas (IAA-CSIC). This study makes use of VLBA data from the VLBA-BU Blazar Monitoring Program (BEAM-ME and VLBA-BU-BLAZAR; http://www.bu.edu/blazars/BEAM-ME.html), funded by NASA through the Fermi Guest Investigator Program. The VLBA is an instrument of the National Radio Astronomy Observatory. The National Radio Astronomy Observatory is a facility of the National Science Foundation operated by Associated Universities, Inc. This study used observations conducted with the 1.8 m Perkins Telescope Observatory (PTO) in Arizona (USA), which is owned and operated by Boston University. The Liverpool Telescope is operated on the island of La Palma by Liverpool John Moores University in the Spanish Observatorio del Roque de los Muchachos of the Instituto de Astrofisica de Canarias with financial support from the UK Science and Technology Facilities Council.
C.C., D.B., and D.A. acknowledge support from the European Research Council (ERC) under the Horizon ERC Grants 2021 programme under the grant agreement No. 101040021. The IAA-CSIC co-authors, C.C., and D.B., acknowledge financial support from the Spanish "Ministerio de Ciencia e Innovaci\'{o}n" (MCIN/AEI/ 10.13039/501100011033) through the Center of Excellence Severo Ochoa award for the Instituto de Astrof\'{i}sica de Andaluc\'{i}a-CSIC (CEX2021-001131-S), and through grants PID2019-107847RB-C44 and PID2022-139117NB-C44. The research at Boston University was supported in part by National Science Foundation grant AST-2108622 and by NASA Fermi Guest Investigator grants 80NSSC23K1507 and 80NSSC23K1508 (and their predecessors. HZ is supported by NASA under award number 80GSFC24M0006 and also IXPE GO program Cycle 1, grant numbers 80NSSC24K1160 and 80NSSC24K1173.
\end{acknowledgments}

\begin{contribution}
%%This section gives authors the space to recognize author contributions. The text inside this environment is NOT counted towards the total word quanta. At a minimum, manuscripts are expected to include this text:

%All authors contributed equally to the Terra Mater collaboration.

%% But authors are expected to provide more specific details, e.g. 
%%
%%SC was responsible for writing and submitting the manuscript.
%%WWM came up with the initial research concept and edited the manuscript.
%%OTS obtained the funding and edited the manuscript.
%%EBF provided the formal analysis and validation. He also edited the manuscript.
%%GEH Supervised the undergraduates, wrote the software and administers the project github and Zenodo repositories.
%%
%% Authors can use the Contributor Role Taxonomy (CRediT) at
%% for ideas on how to write a good statement tailored to their needs.

CC and DB were responsible for writing the manuscript and analyzing the data. IA conceived the research goals and led the POLAMI program. IA, CC, IM, CT, JEP, and DAO were responsible for proposal writing, observations, and data calibration for the POLAMI program. SJ, AM, MJ, and ZRW were responsible for data acquisition from the Perkins Telescope and contributed to the scientific discussion underlying the study. HZ contributed to the scientific discussion and edited the manuscript. CMC, HJ, and IAS were responsible for the acquisition and analysis of data from the Liverpool Telescope. GAB, TSG, EGL, DAM, SSS, IST, YVT, and AAV were responsible for the acquisition, calibration, and analysis of data from the Crimean Astrophysical Observatory and the LX-200 Telescope of St. Petersburg University. We thank the anonymous referee for the positive review.
\end{contribution}

\bibliography{bibliography}{}
\bibliographystyle{aasjournalv7}

\appendix %First appendix
\section{EVPA distribution in sample sources.}
\label{app:A}

%%%%% Figure %%%%%
\begin{figure*}[h!]
   \includegraphics[width=.30\textwidth]{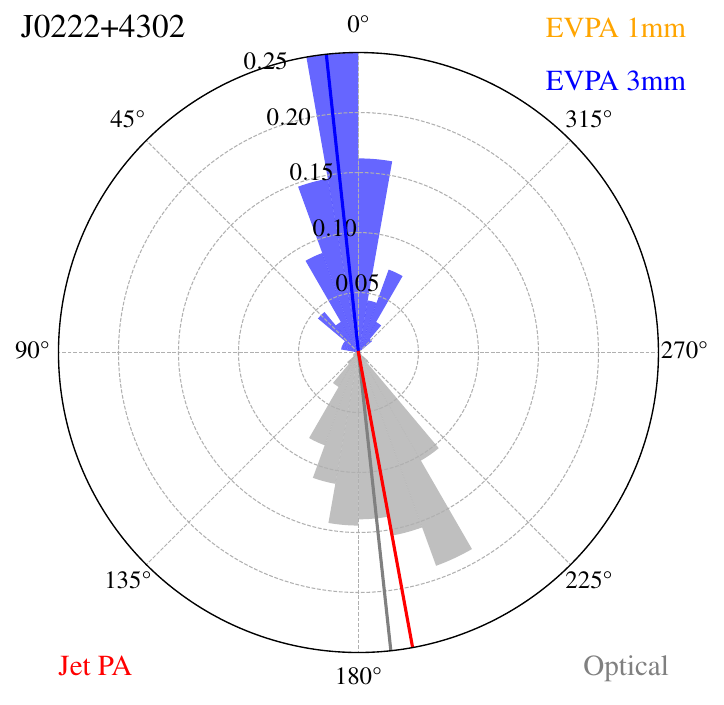}
   \includegraphics[width=.30\textwidth]{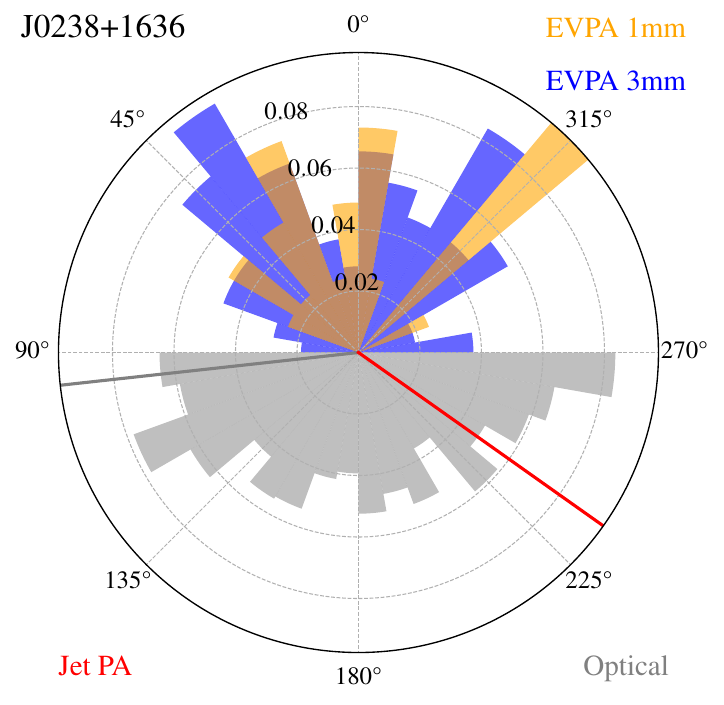}
   \includegraphics[width=.30\textwidth]{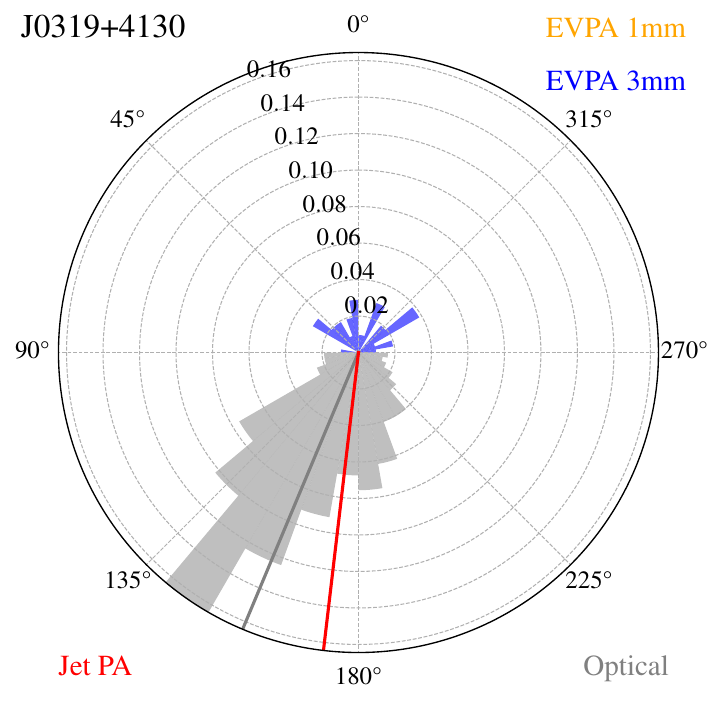}\\
   \includegraphics[width=.30\textwidth]{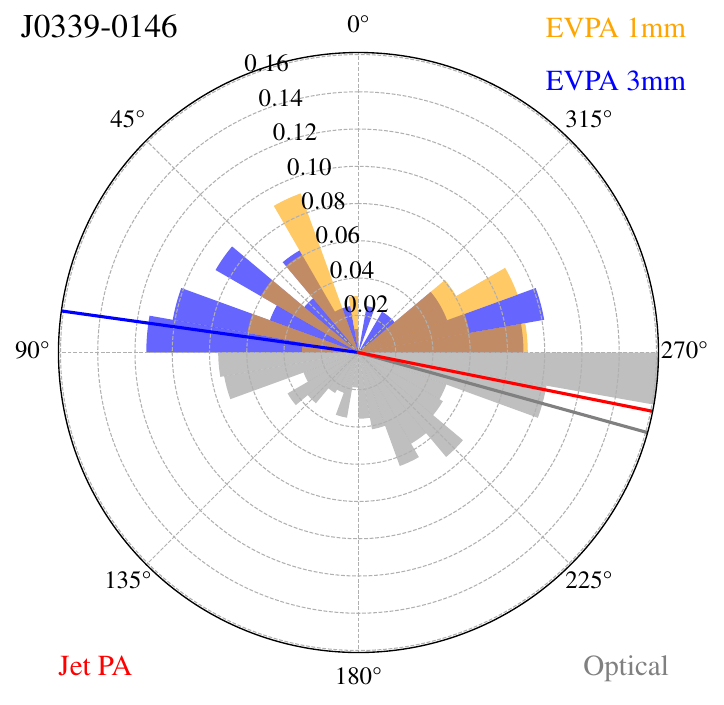}
   \includegraphics[width=.30\textwidth]{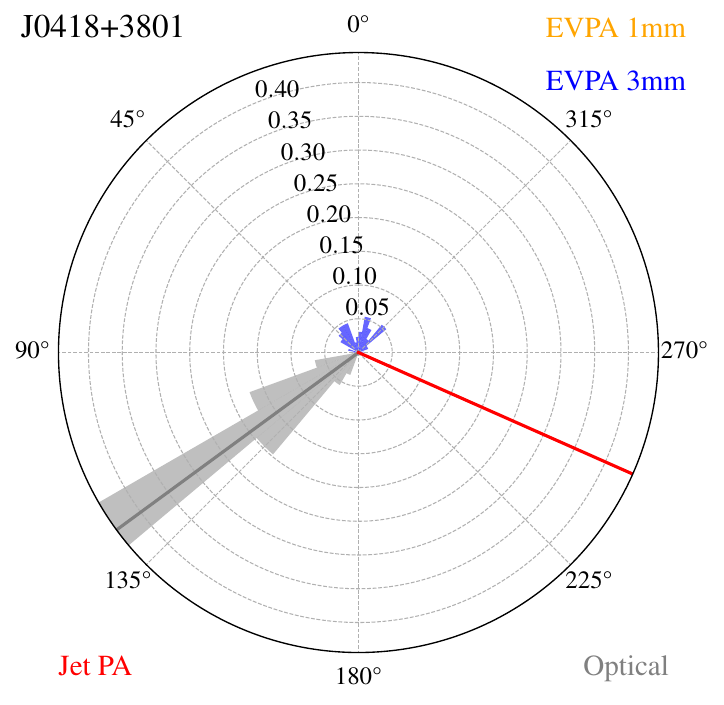}
   \includegraphics[width=.30\textwidth]{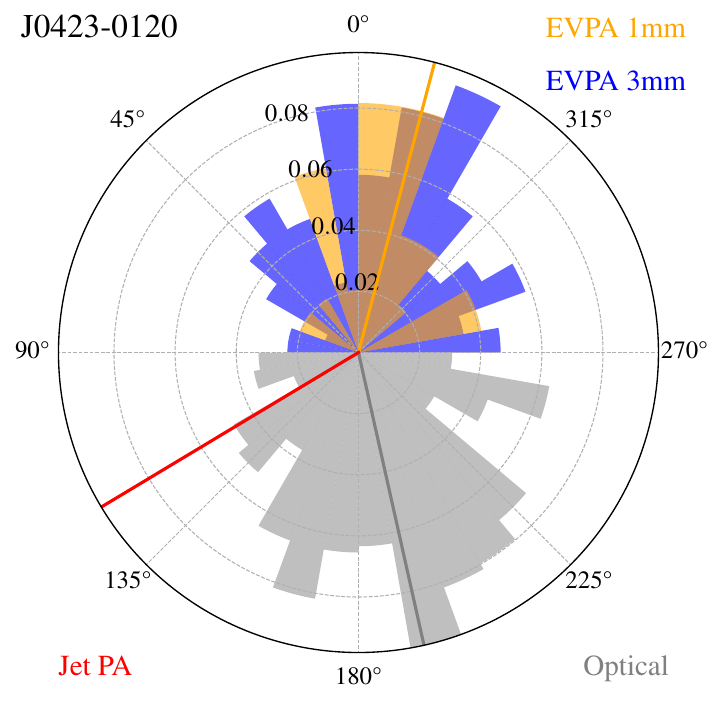}\\
   \includegraphics[width=.30\textwidth]{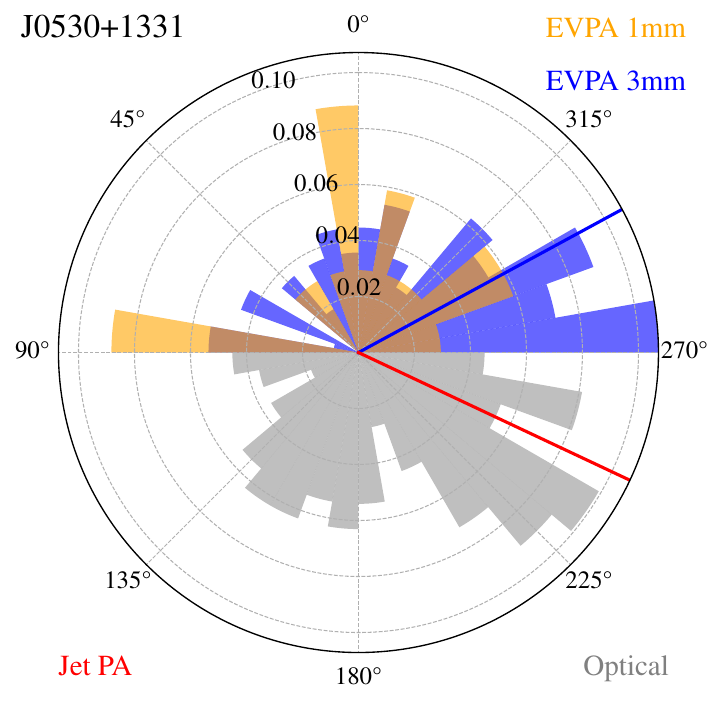}
   \includegraphics[width=.30\textwidth]{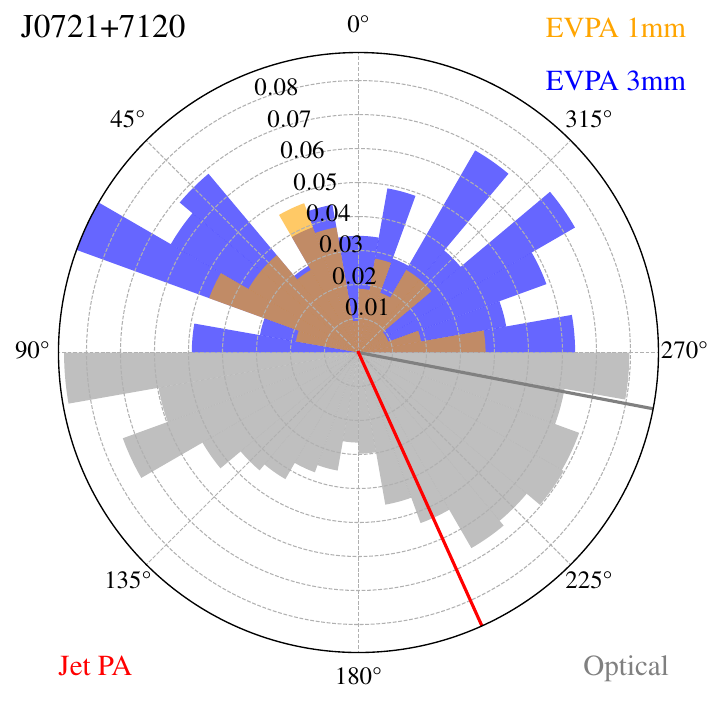}
   \includegraphics[width=.30\textwidth]{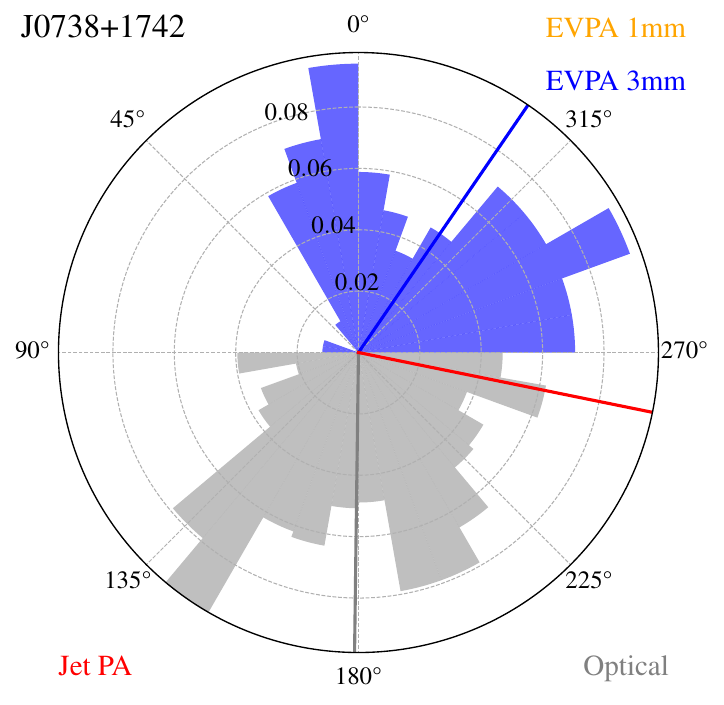}
\caption{Millimeter (upper hemisphere) and optical (lower hemisphere) EVPA distribution in AGN of the monitoring sample. EVPA distributions at 3~mm (blue), 1~mm (yellow), and optical (grey) wavelengths are overlaid by vectors of the same color representing the mean value of the respective distribution. The red line marks the jet PA. The different circles mark the counts as a fraction of the total.}
\label{fig:evpa1}
% ./polami/5_circular_hist/plot_hist_opt_mm.py
\end{figure*}

\begin{figure*}
\ContinuedFloat
    \includegraphics[width=.30\textwidth]{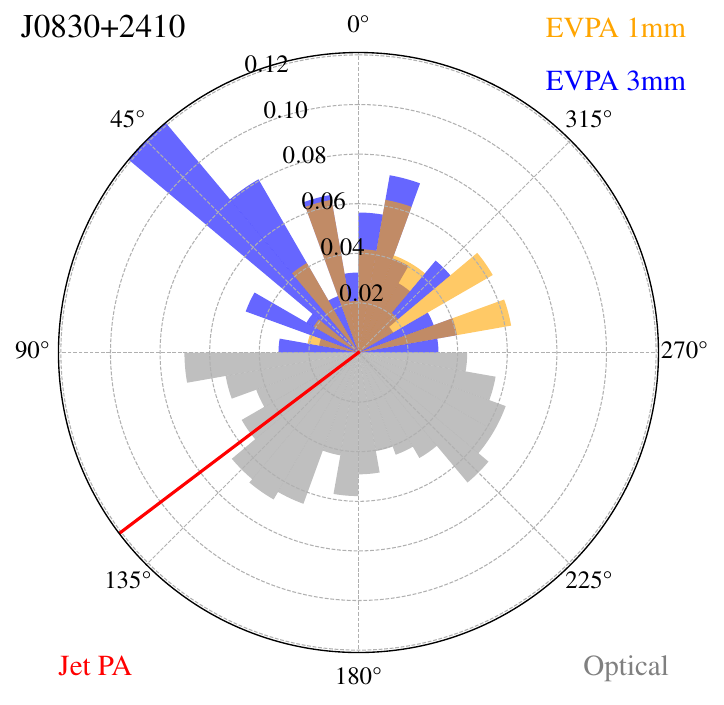}
    \includegraphics[width=.30\textwidth]{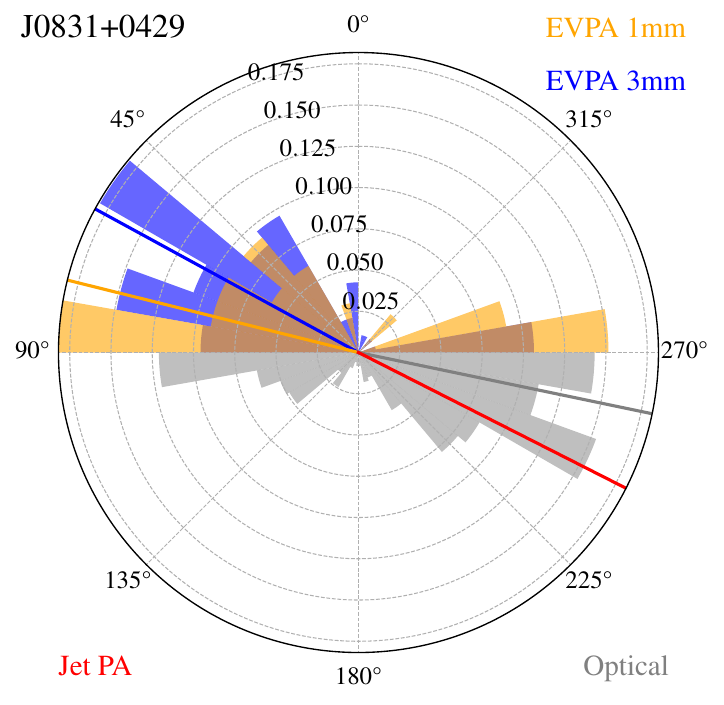}
    \includegraphics[width=.30\textwidth]{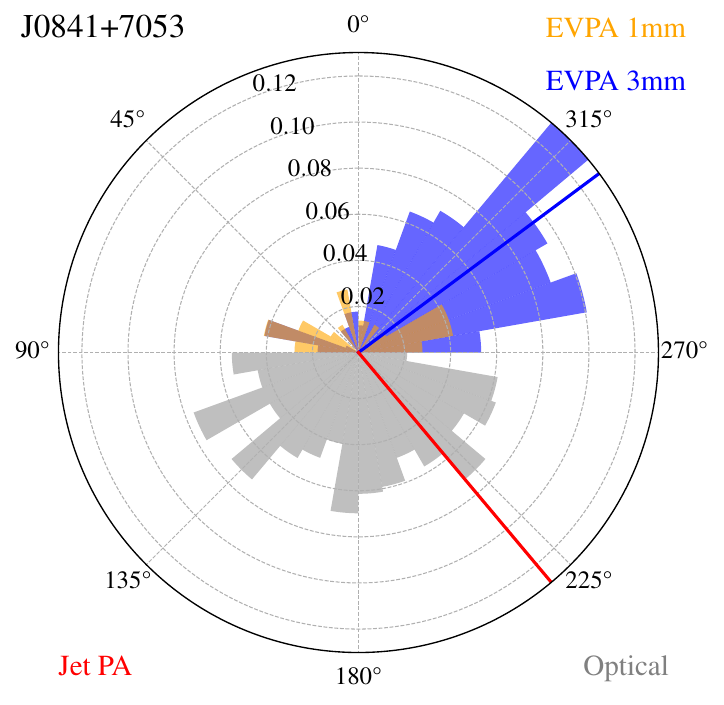}\\
    \includegraphics[width=.30\textwidth]{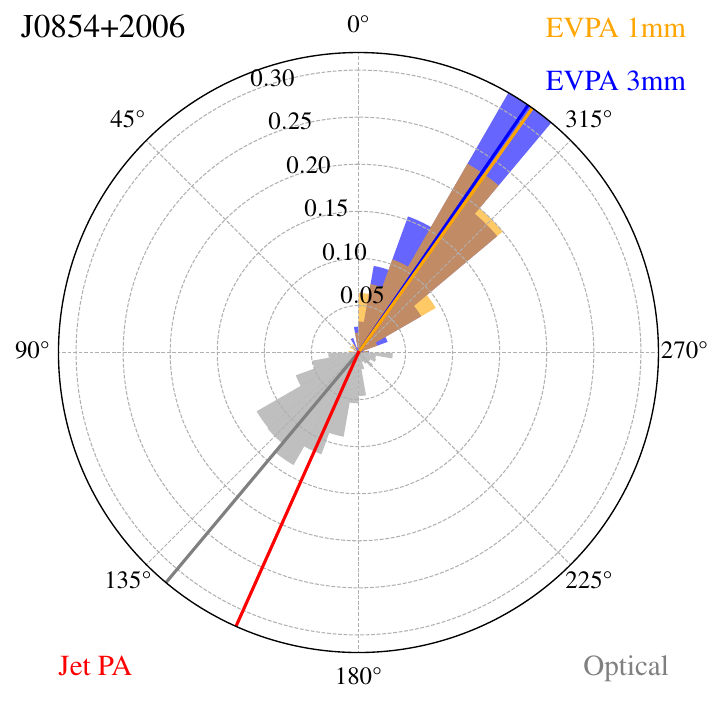}
    \includegraphics[width=.30\textwidth]{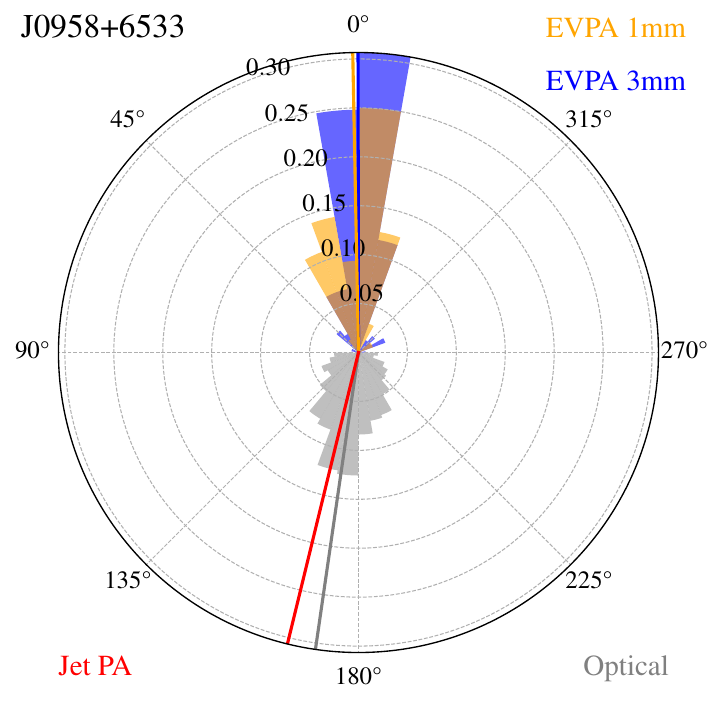}
    \includegraphics[width=.30\textwidth]{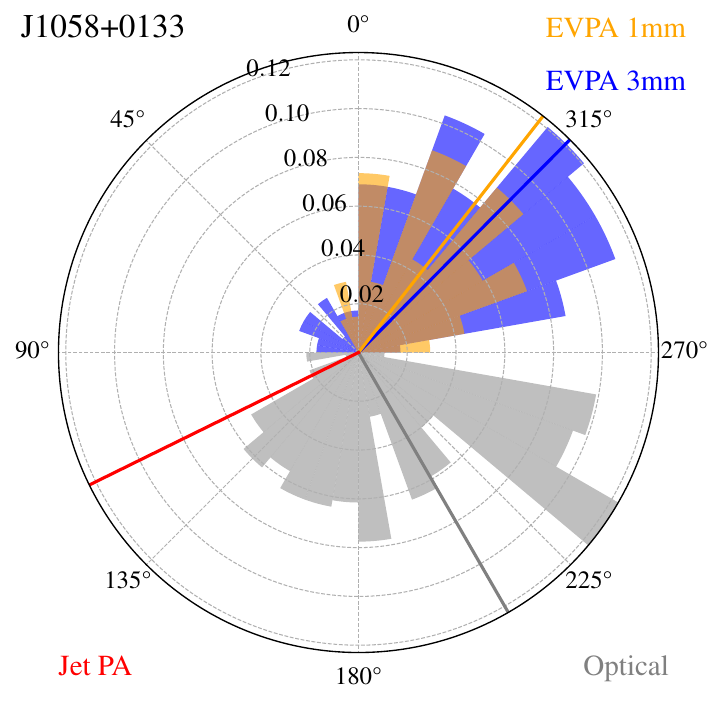}\\
    \includegraphics[width=.30\textwidth]{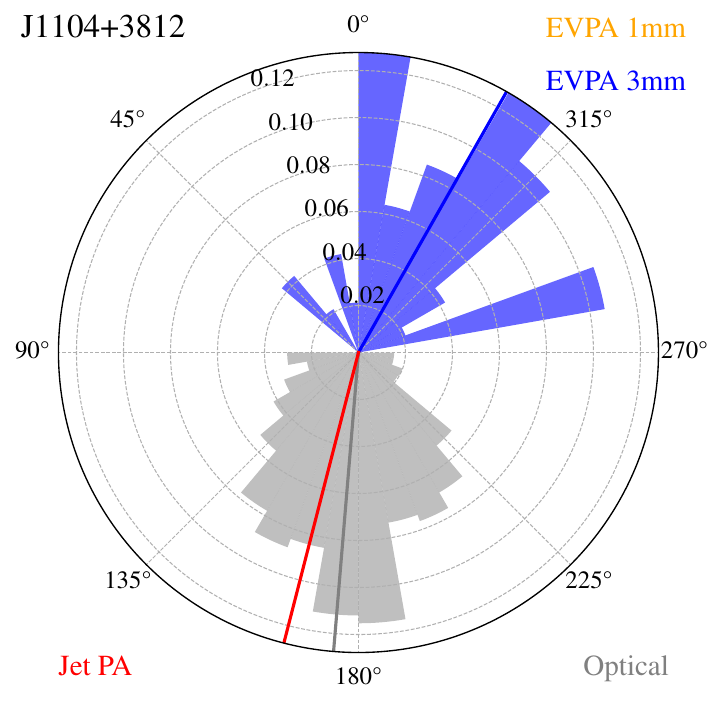}
    \includegraphics[width=.30\textwidth]{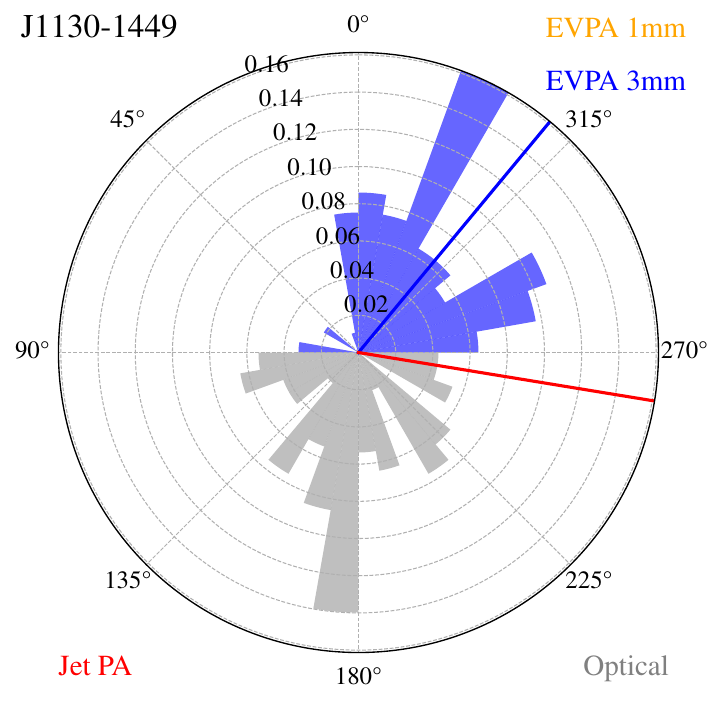}
    \includegraphics[width=.30\textwidth]{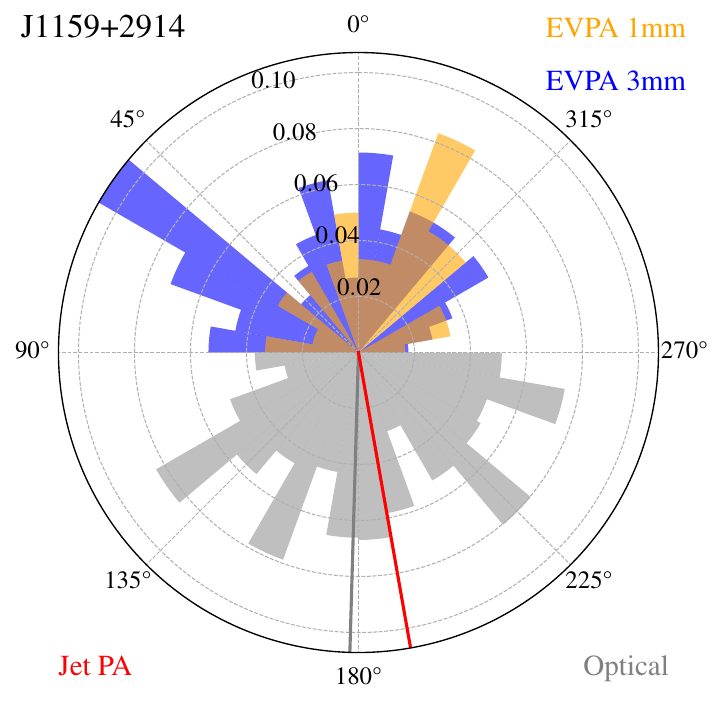}\\
    \includegraphics[width=.30\textwidth]{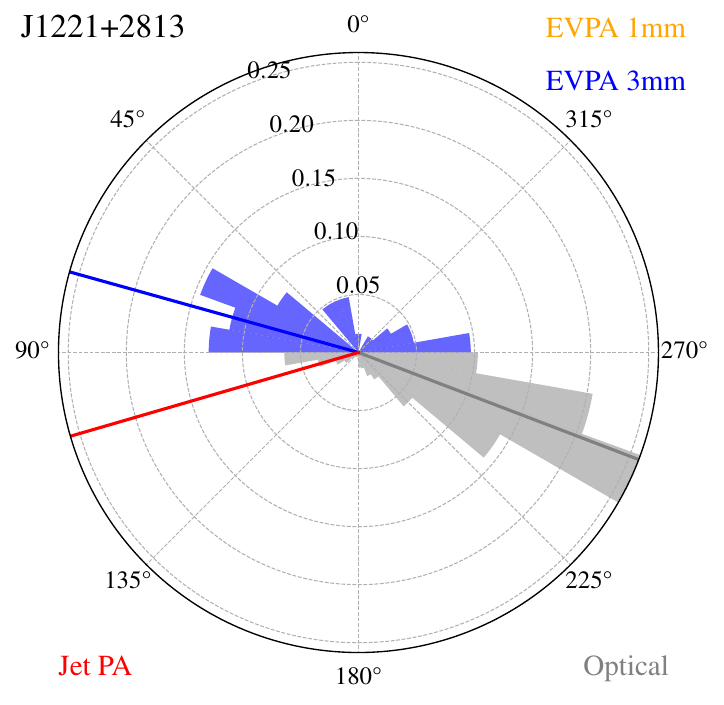}
    \includegraphics[width=.30\textwidth]{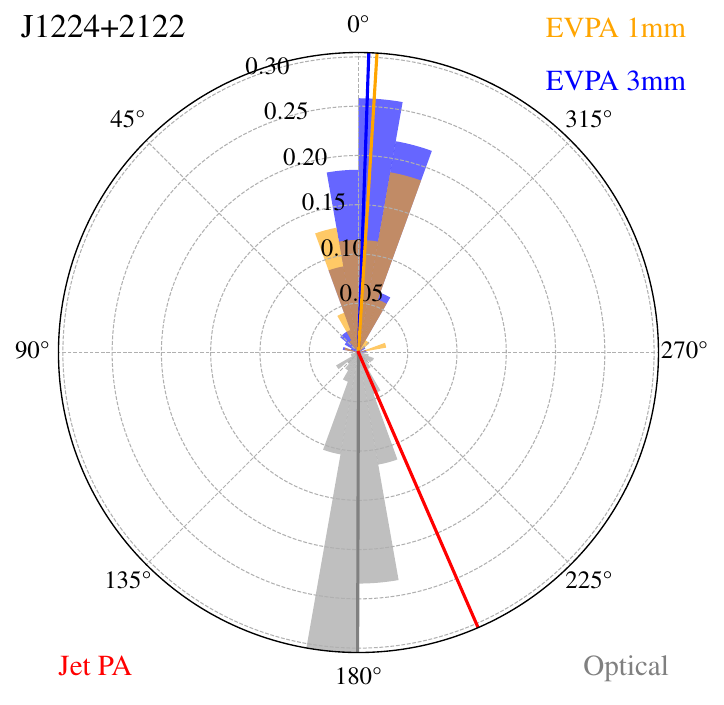}
    \includegraphics[width=.30\textwidth]{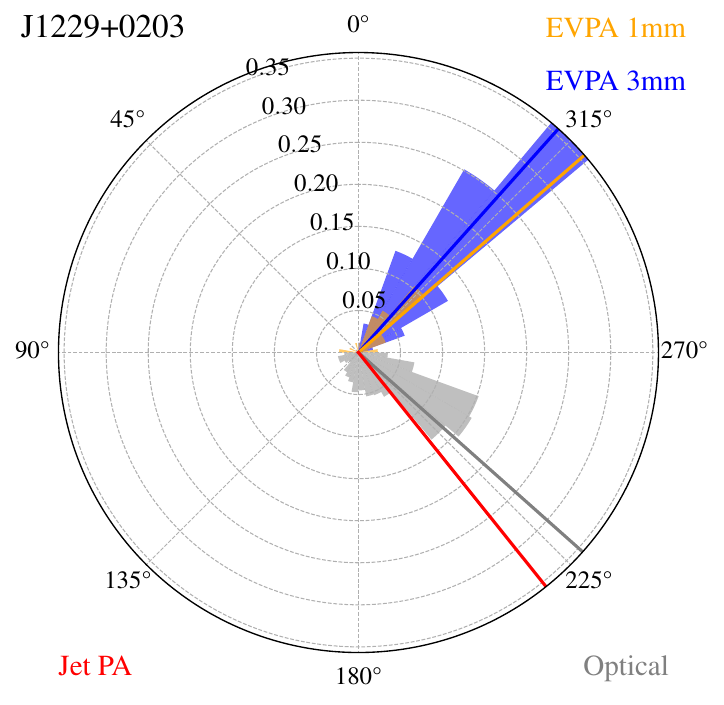}
\caption{{\it Continued}}
% ./polami/5_circular_hist/plot_hist_opt_mm.py
\end{figure*}

\begin{figure*}
\ContinuedFloat
    \includegraphics[width=.30\textwidth]{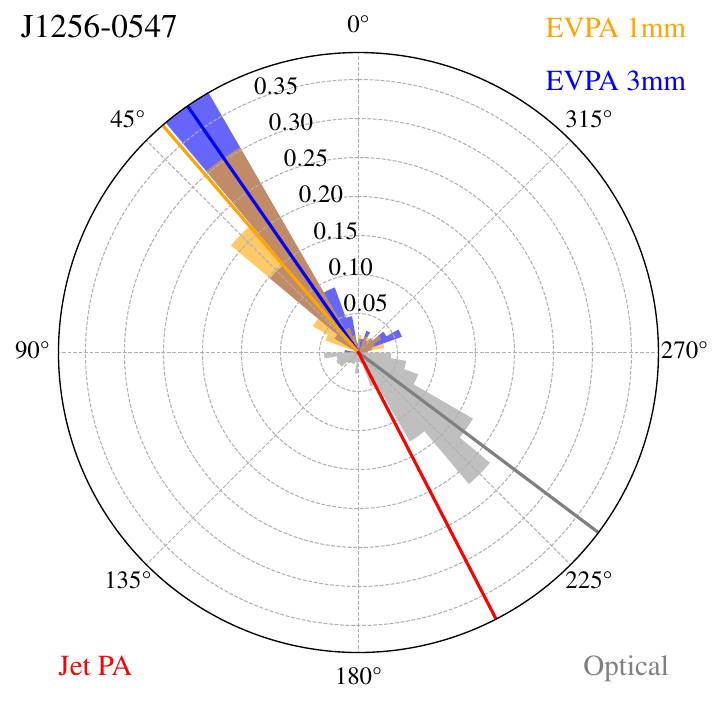}
    \includegraphics[width=.30\textwidth]{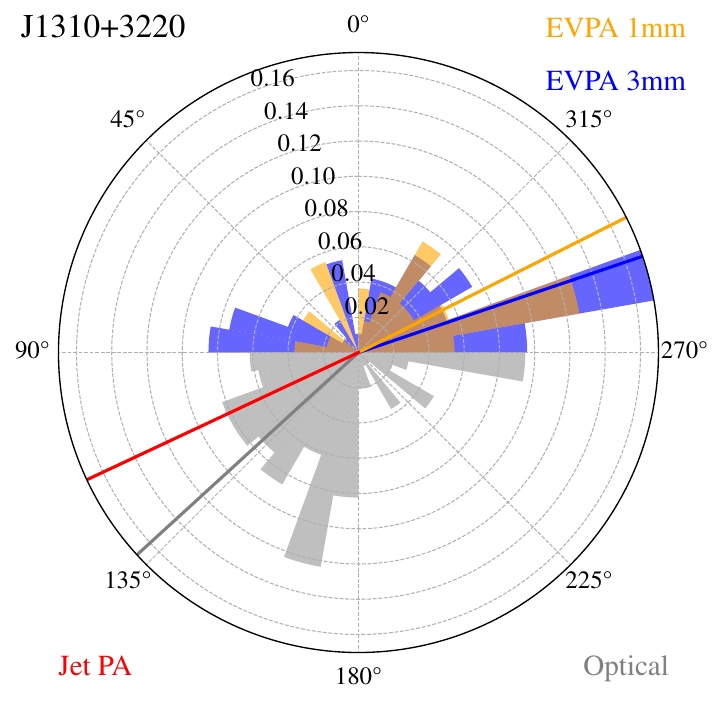}
    \includegraphics[width=.30\textwidth]{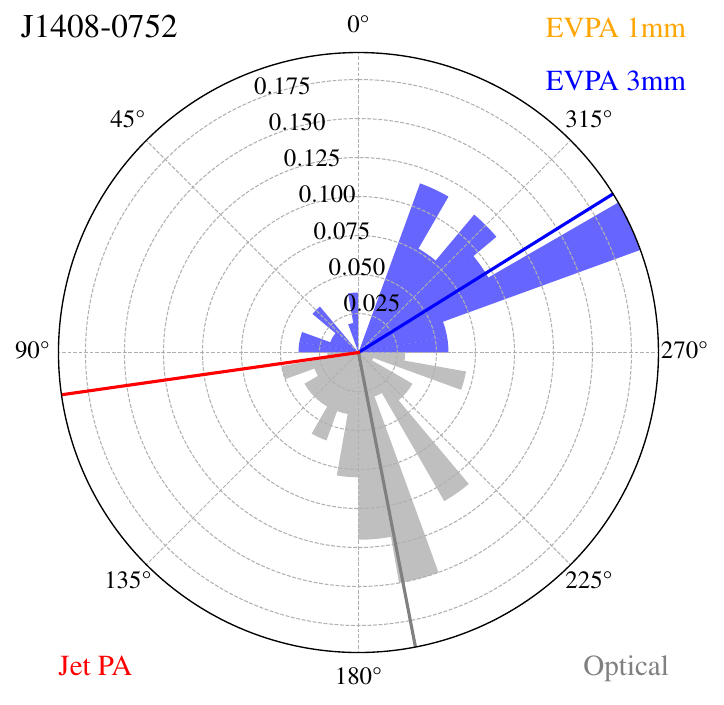}\\
    \includegraphics[width=.30\textwidth]{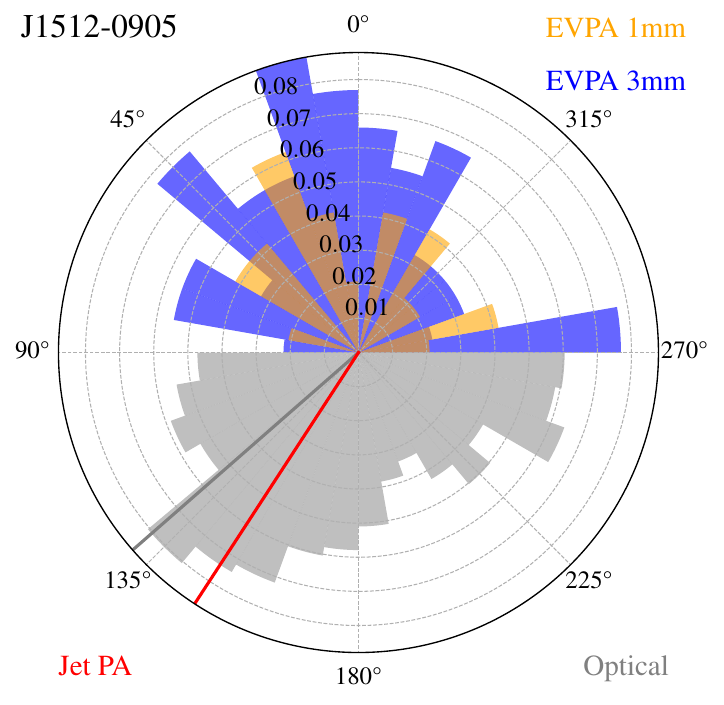}
    \includegraphics[width=.30\textwidth]{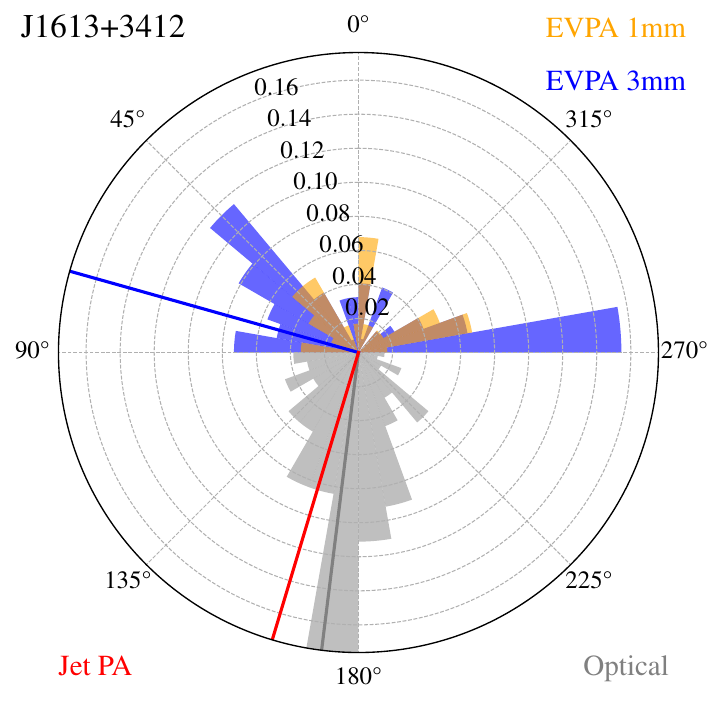}
    \includegraphics[width=.30\textwidth]{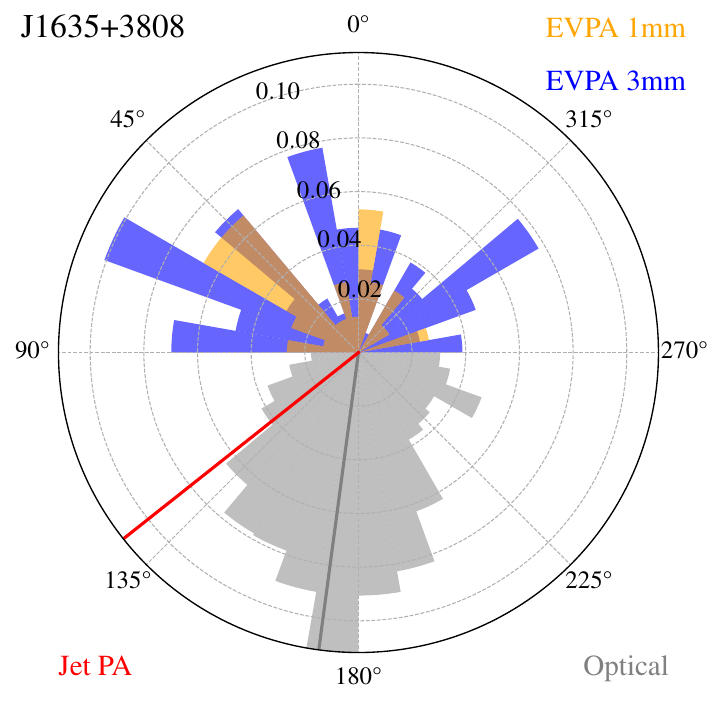}\\
    \includegraphics[width=.30\textwidth]{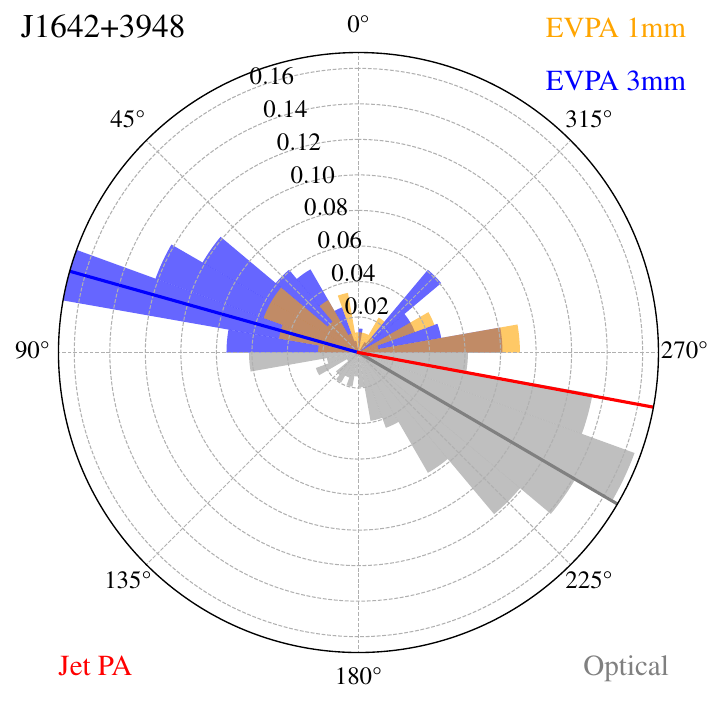}
    \includegraphics[width=.30\textwidth]{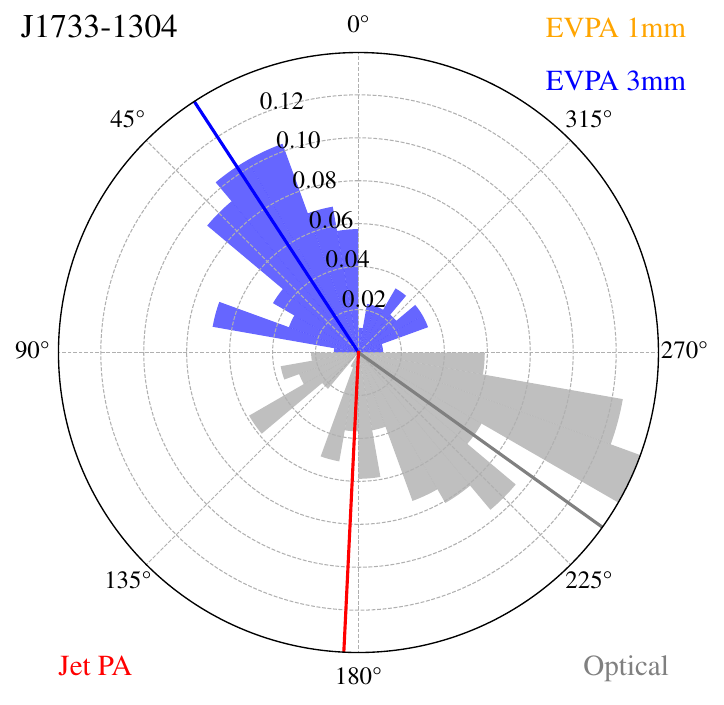}
    \includegraphics[width=.30\textwidth]{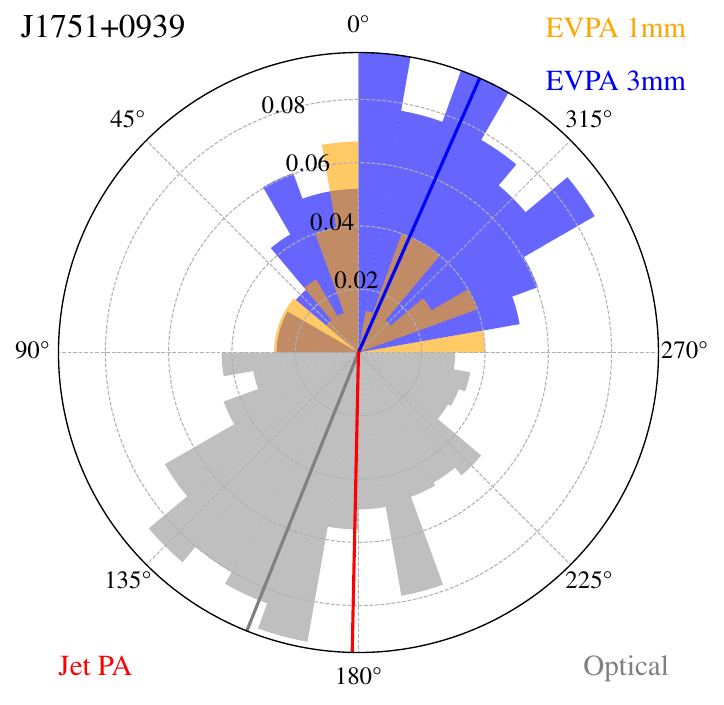}\\
    \includegraphics[width=.30\textwidth]{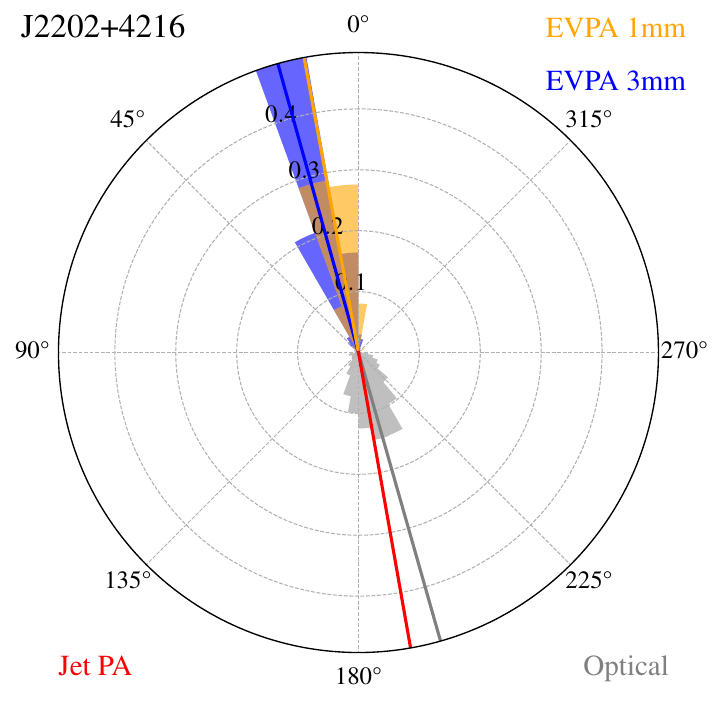}
    \includegraphics[width=.30\textwidth]{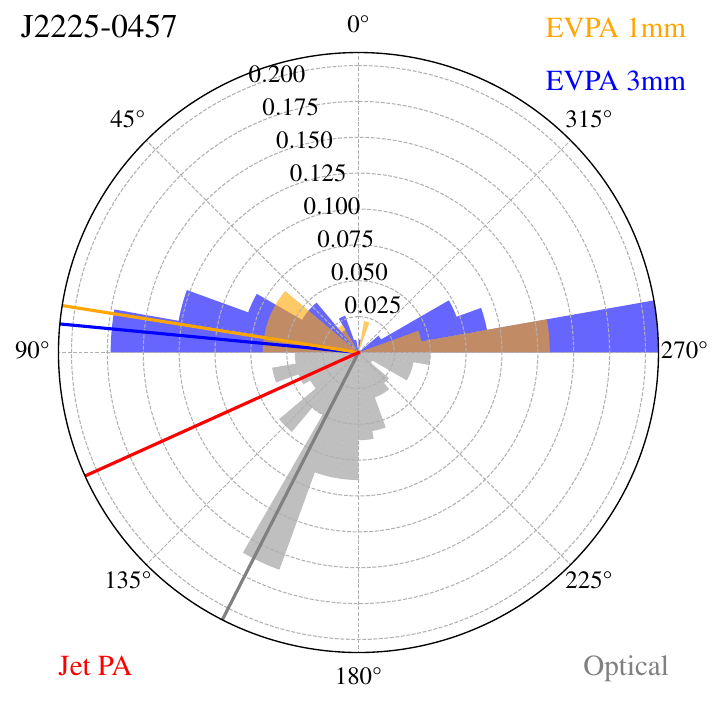}
    \includegraphics[width=.30\textwidth]{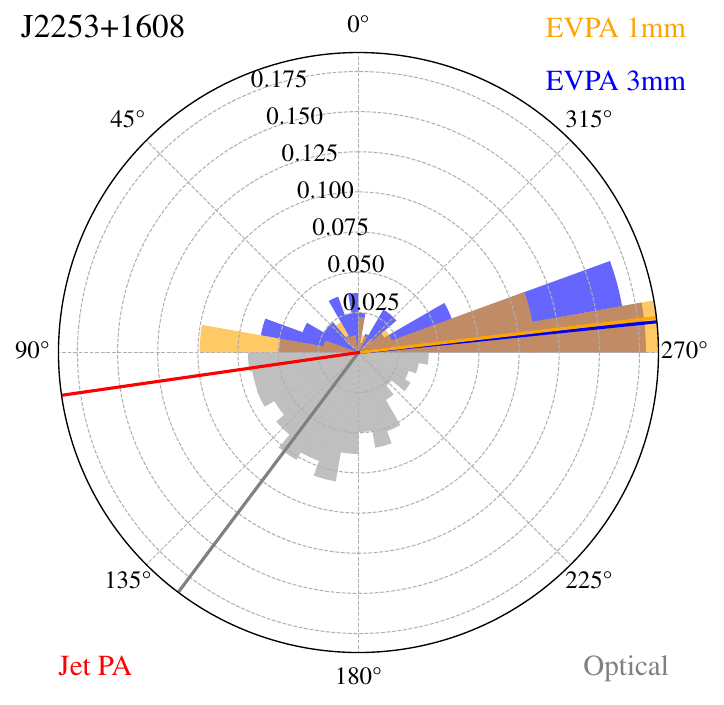}
\caption{{\it Continued}}
% ./polami/5_circular_hist/plot_hist_opt_mm.py
\end{figure*}



\end{document}